%% file: ms.tex
\documentclass[%
  fontsize=10pt,%
  paper=a4,%
  bibliography=totoc,%
  headsepline=true,%
]{scrartcl}
\pdfoutput=1

\KOMAoption{abstract}{true}
\setkomafont{section}{\large}
\setkomafont{subsection}{\normalsize}
\setkomafont{subsubsection}{\normalsize}
\setkomafont{paragraph}{\normalsize}

\usepackage[automark]{scrlayer-scrpage}
  \clearpairofpagestyles
  \ohead[]{\pagemark}
  \ihead[]{\headmark}
  \ofoot[\pagemark]{}
\usepackage[inner=2.5cm, outer=2.5cm, top=3cm, bottom=3cm]{geometry} 
\usepackage[T1]{fontenc}
\usepackage[utf8]{inputenc}
\usepackage{lmodern}          
\usepackage[british]{babel} 
\usepackage[colorlinks=true]{hyperref}
\usepackage{csquotes}
\usepackage[%
  style=alphabetic,%
  doi=true,%
  url=true,%
  backend=biber,%
  giveninits=true,%
  sorting=nyt,%
  maxbibnames=6%
]{biblatex}
  \bibliography{files/Literatur}

\usepackage{amssymb,amsmath,amsthm}
\usepackage{mathtools,units}
\usepackage{commath}      
\usepackage{nicefrac}
\usepackage{array}  
\usepackage{multirow}
\usepackage{bm}
\usepackage{dsfont}
\usepackage{calrsfs}
\usepackage{stmaryrd}
\DeclareMathAlphabet{\pazocal}{OMS}{zplm}{m}{n}
\DeclareMathAlphabet{\pazocalbf}{OMS}{cmsy}{b}{n}
\usepackage{etoolbox}
\preto\subequations{\ifhmode\unskip\fi}
                
\usepackage{enumitem}
  \setlist[description]{
    font={\mdseries\rmfamily},
    labelindent=\parindent,
    leftmargin=!
    }
  \setlist[itemize]{nosep, topsep=0pt, wide = 1em, leftmargin=*} 
\usepackage[section]{placeins}
\usepackage{booktabs}
\usepackage{multirow}
\usepackage{siunitx}
\usepackage{subfig}
\usepackage{graphicx} 
\usepackage[export]{adjustbox}
\usepackage[dvipsnames]{xcolor}
\usepackage{tikz}
\usepackage{pgfplots}
  \pgfplotsset{compat=1.16}

\input{files/definitions.tex}

\begin{document}

\title{\Large Falling balls in a viscous fluid with contact: Comparing
numerical simulations with experimental data}
\author{
\scshape\normalsize
Henry von Wahl 
\thanks{Corresponding author:
\texttt{henry.vonwahl@ovgu.de}}
\textsuperscript{,}%
\thanks{Insitut für Analysis und Numerik, Otto-von-Guericke Universität,
Universitätsplatz 2, {D-39106} Magdeburg}
\and
\scshape\normalsize
Thomas Richter\footnotemark[2] 
\and
\scshape\normalsize
Stefan Frei
\thanks{Department of Mathematics and Statistics, University of Konstanz,
Universitätsstraße 10, {D-78457} Konstanz}
\and
\scshape\normalsize
Thomas Hagemeier
\thanks{Institut für Strömungstechnik und Thermodynamik, Otto-von-Guericke
Universität, Universitätsplatz 2, {D-39106} Magdeburg}
}

\date{\small\today}
\maketitle

\begin{abstract}
We evaluate a number of different finite element approaches for fluid-structure
(contact) interaction problems against data from physical experiments.
For this we take the data from experiments by Hagemeier [Mendeley Data,
\textsc{doi}: 
\href{https://doi.org/10.17632/mf27c92nc3.1}{\texttt{10.17632/mf27c92nc3.1}}].
This consists of trajectories of single particles falling through a highly
viscous fluid and rebounding off the bottom fluid tank wall. The resulting flow
is in the transitional regime
between creeping and turbulent flows. This type of configuration is
particularly challenging for numerical methods due to the large change of the
fluid domain and the contact between the wall and particle. In the numerical
simulations we consider both rigid body and linear elasticity models for the
falling particles. In the first case, we compare results obtained with the well
established Arbitrary
Lagrangian Eulerian (ALE) approach and a moving domain CutFEM method together
with
a simple and common approach for contact avoidance. 
For the full fluid-structure interaction (FSI) problem with contact, we use
a fully Eulerian approach in combination with a unified FSI-contact treatment
using Nitsche's method.
For higher computational efficiency we use the geometrical symmetry of the
experimental set up to reformulate the FSI system into two spatial dimensions.
Finally, we show full three dimensional ALE
computations to study the effects of small perturbations in the initial state
of the
particle to investigate deviations from a perfectly vertical fall observed in
the experiment.
The methods are implemented in open-source finite element libraries and the
results are made freely available to aide reproducibility.
\end{abstract}

\input{files/intro}

\input{files/description}

\input{files/reduced_model}

\input{files/numerical_experiment}

\input{files/conclusion}

\section*{Acknowledgements}
HvW and TR acknowledge support by the Deutsche Forschungsgemeinschaft (DFG,
German Research Foundation) - 314838170, GRK 2297 MathCoRe. TR further
acknowledges support by the Federal Ministry of Education and Research of
Germany (project number 05M16NMA).

\renewcommand{\bibfont}{\normalfont\footnotesize}
\setlength\bibitemsep{.33\itemsep}
\printbibliography

\clearpage
\appendix
\input{files/appendix}

\end{document}

%% file: files/intro.tex
\section{Introduction}

Flows containing particles, i.e., particulate flows or particles settling in a
fluid, have many industrial and biological applications. Examples range from
the transport of platelets in blood flows to the sedimentation of sand in
pipelines.

We shall consider single elastic spherical particles falling freely in a
viscous fluid and rebounding off the bottom wall of the fluid domain at
Reynolds numbers in the transitional regime between creeping and turbulent
flows. The multiphase and fluid-structure interaction (FSI) problem with solid
contact posed by the settling in the fluid and rebounding off a wall is
challenging from both an analytical and numerical perspective.

From the theoretical point of view, the correct model for the transition to
contact with the bottom wall is not yet fully understood. In the case where a
rigid solid is assumed, most flow models lead to
results contradicting real world observations. For example, if a creeping flow
is assumed, such that the linear Stokes equations are applicable, then contact
can only occur under singular forces, c.f.~\cite{Bre61}. When the non-linear
incompressible Navier-Stokes together with no-slip boundary conditions are
taken for the fluid model, then contact cannot occur and it is impossible to
release contact~\cite{Fei03a}. This can however be overcome, if the boundary
condition is modified to a free-slip condition~\cite{GHW15}, the rough nature
of the surface is taken into account~\cite{GH14} or the fluid is taken to be
compressible~\cite{Fei03b}. If the solid model is changed to take the
elasticity of the body into account, then it is currently assumed, that even
with perfectly smooth boundaries and incompressibility, rebounding without
contact can occur due to the storage of energy in the elastic 
solid~\cite{DSH86, GH16, HT09}. This has been refined recently (in the Stokes
setting)~\cite{GSST20}, where it has been shown, that internal storage of
energy is not sufficient, but that additionally a change in the
``flatness'' is necessary to achieve physically meaningful rebound without
topological contact.

For numerical methods, the challenge lies in the discretisation of the
resultant FSI system~\cite{Richter2017}. It consists of a free boundary value
problem with a moving interface. The most well established method for this is
the Arbitrary Lagrangian Eulerian (ALE) approach~\cite{DGH82}. This approach
leads to very efficient and accurate computations in situations where the 
method is usable. However, its usage limited as it breaks down when
deformations with respect to the reference configuration become too large and
when contact occurs~\cite{FreiRichterWick2016}.
To deal with large deformations, overlapping mesh techniques have been
developed~\cite{JLL15}. Here the background fluid domain
and the region around the structure are meshed separately so that the
fluid-solid interface is resolved. The two meshes are then coupled using
unfitted approaches. This then allows a hybrid approach, where the solid and
the near fluid are treated using the ALE framework, while the remaining fluid
is treated in Eulerian coordinates~\cite{SAW19a}. To overcome both large
deformations and contact, fully Eulerian approaches have lately become the
focus of research. 
In the case of rigid bodies, a number of different approaches have
been considered. For example, based on fictitious domain methods using Lagrange
multipliers~\cite{GPHJ99}, XFEM type approaches~\cite{CFL13, CF15} and most
recently CutFEM approaches using Nitsche's method~\cite{BFM19, WRL20}.
Here a major issue remains that of achieving a realistic rebound
effect, since an artificial contact/lubrication force is added to the
equation governing the motion of the solid to prevent overlap of the solid 
regions~\cite{GPHJ99}. Nevertheless, topological changes appear to be
unproblematic for the CutFEM type approaches~\cite{LO19}. 

Considering full fluid-structure interactions, immersed approaches have become
popular in recent years~\cite{HansboHermansson2003, LegayChessaBelytschko2006,
GerstenbergerWall2008, BurmanFernandez2014}. Here the fluid
and the solid are treated in their natural Eulerian and Lagrangian coordinate
systems, respectively, and the sub-domains are meshed separately.
The two meshes are then coupled by means of Nitsche's 
method~\cite{BurmanFernandez2014, HansboHermansson2003} or using Lagrange
multipliers~\cite{Baaijens2001, LegayChessaBelytschko2006,
GerstenbergerWall2008}.
Another possibility to handle large deformations and contact are fully Eulerian
approaches, where both the solid and fluid equations are formulated in the
Eulerian coordinate framework, which simplifies the coupling within monolithic
algorithms~\cite{Dunne2006, Cottetetal2008, Richter2012b, Fre16,
HechtPironneau2017}. All these approaches are however relatively new and
require further development with respect to accuracy and robustness.

The aforementioned methods have been applied to different test cases for
numerical validation and a priori error estimates are also available in most
cases. Established benchmarks for fluid structure interaction
problems such as~\cite{TH06} completely avoid contact, since the methods which
handle contact remain relatively new. For rigid body motion,
most numerical studies are interpreted qualitatively or compared to artificial,
analytically derived solutions. Especially in the cases where artificial forces
are introduced in order to avoid contact of rigid solids, real validation is
near impossible, as this introduces model parameters for which there is no a
priori knowledge on a good choice. However, a number of FSI methods have
recently become available that are able to resolve 
contact~\cite{AgerWalletal, ASW19, BurmanFernandezFreiGerosa19,
BurmanFernandezFrei2020, ZAV20}. This then raises
the question of how well the different modelling and discretisation approaches
depict the behaviour of contact and rebound observed in physical experiments.

In this work, we take recently published data from experiments where
different solid spherical particles were allowed to settle in a viscous 
fluid~\cite{HTR20, Hag20}. We then use a rigid-body ALE, a rigid-body Eulerian
CutFEM and a fully Eulerian FSI approach to each simulate the scenarios
presented by the physical experiments. 
This aims to show the validity and the applicability of these
different approaches to the different aspects/problems posed by this process.
Furthermore, we will illustrate how spatially reduced models are able to
capture the behaviour in comparison to full three dimensional computations.
To the best of our knowledge, there is currently no comparable benchmark which
considers such a multiphase flow/ fluid structure interaction problem with
contact which is validated against experimental data.

The remainder of this paper is structured as follows. 
In \autoref{sec:description} we describe
the considered problem, that is a description of the physical experiment, the
mathematical models used to described the experiment, the specific
set-ups we will simulate and the quantities used to compare the numerical
simulations with the experimental data. 
\hyperref[sec:reduced-model]{Section~\ref*{sec:reduced-model}} then
briefly covers the reduced formulation we apply to increase the computational
efficiency in our numerical methods. The numerical computations are then
presented in \autoref{sec:num-comp}; we present the details of the different
numerical approaches in \autoref{subsec:discretisations} and the results are
then presented in \autoref{subsec:results}. We discuss the conclusions from
these results in \autoref{sec:conclusion} and consider the aspect which remain
open.
Furthermore, we define and compute two simplified set-ups in 
\autoref{appendix}, designed to help others reproduce the presented
computational results.

%% file: files/description.tex
\section{Description}\label{sec:description}

We describe the experimental set-up used to gather the data, the
mathematical model we will use to reproduce the behaviour observed in the
experiments and we define relevant quantities used to compare the results
quantitatively.

\subsection{Physical Experiment}\label{subsec:experiment}

The experiments in~\cite{HTR20} capture the settling and impact process of
spherical particles with different size and density in a cylindrical tank. The
later contains a liquid mixture consisting of glycerine and water at equal
volume fractions. 
At the bottom of the cylindrical tank, which is made of acrylic glass for
optical access, a massive steel anvil serves as impact object.
Moreover, the cylinder is surrounded by a rectangular container, filled with
refractive index matching liquid, to compensate for optical distortion coming
from the curved cylinder walls. The filling level allows the observation of the
particle settling along a vertical distance of $140$ to
$160~\si{\milli\metre}$, depending on the particle size. Initially, each
particle is held in place and submerged in the liquid by a vacuum tweezer. The
particle is then released by switching off the vacuum pump. The particle is
tracked during the settling process and the impact on the steel anvil,
including the rebound using a high-speed CMOS-camera. This acquires shadow
images at a frame rate of 1000 frames per second and a scale factor of 8.89
pixel per mm. An image processing algorithm coded in MATLAB yields the in-plane
particle coordinates as function of time and allows to extract the
instantaneous particle settling velocity.  

The resulting data is available via Mendeley Data~\cite{Hag20}. This data is
the basis for our comparison and validation of the numerical code.

\subsection{Mathematical model}\label{subsec:model}

We consider a bounded domain $\O\in\R^d$, with $d\in\{2,3\}$,  over a finite,
non-empty, time interval $[0,T_{end}]$. This is divided into a $d$-dimensional
fluid region $\FL$, a $d$-dimensional solid $\SO$ and a $d-1$-dimensional
interface $\IN$ dividing the solid and fluid regions. For these we have
$\O=\FL\:\dot\cup\:\IN\:\dot\cup\:\SO$.

\paragraph{Fluid model} In the time dependent fluid region $\FL(t)$, we
consider the incompressible Navier-Stokes equations. Find a velocity $\u$ and a
pressure $p$ such that
\begin{subequations}\label{eqn:Navier-Stokes}
\begin{align}
  \rho_f\left(\partial_t\u + (\u\cdot\nabla)\u\right) - \div\stress
    &=0\label{eqn:Navier-Stokes-MomentumBallance}\\
  \div(\u) &= 0
\end{align} 
\end{subequations}
with the non-symmetric stress tensor
\begin{equation*}
  \stress = \mu_f\nabla\u - p\Id
\end{equation*}
where $\mu_f=\rho_f\nu_f$ is the fluid's dynamic viscosity, $\nu_f$ is the
kinematic viscosity and $\rho_f$ is the fluid's density. Appropriate boundary
conditions to complete this system will be discussed later. Note that we use
bold face letters to denote vector/matrix valued quantities while regular faced
letters denote scalar objects.

\paragraph{Elastic solid and fluid-structure interaction}

We consider a linear elastic solid model in $\SO$ for the solid displacement
$\d$ and the solid velocity $\dot{\d}$, given by
\begin{align}\label{eq:linElast}
\rho_s\partial_t \dot{\d} 
-  \div \sstress\left(\d \right)   
=  (\rho_s-\rho_f) g, \qquad \dot{\d} = \partial_t \d,
\end{align}
with the acceleration due to gravity
$g=\SI[round-mode=off]{-9.807}{\metre\per\second\squared}$
and the Cauchy stress tensor $\sstress$ defined by
\begin{align*}
\sstress(\d) = 2\mu_s E(\d) + \lambda_s \text{tr} (E(\d)) I, 
 \qquad E(\d)=\frac{1}{2} \left( \nabla \d + \nabla \d^T\right)
 \end{align*}
where $\mu_s$ and $\lambda_s$ denote the Lam\'e parameters. Note that we have
subtracted the fluid gravitational force $\rho_f g$ on the right-hand side of
the solid equation to be consistent with the equations for a rigid solid
presented in the following paragraph. Alternatively this could be added on the
right-hand side of the fluid equations.

Solid and fluid are coupled by means of no-slip coupling conditions on $\IN$
\begin{align}\label{eq:FSIcoupling}
\u = \dot{\d}, \qquad \stress \n= \sstress(\d) \n.
\end{align}
The current position of the interface $\IN(t)$ is determined by the
displacement variable $\d$.

\paragraph{Rigid solid} 
As the solid materials we will consider are relatively hard, the consideration
of a rigid solid yields a good approximation of the FSI dynamics, at least up
to the moment when the solid comes close to the lower wall.
The movement of the solid is governed by Newton's second law of motion. Let
$\c_\SO (t)$ be the centre of mass of the solid $\SO$. Since we will consider
spherical particles, this is then governed by
\begin{equation}\label{eqn:Ball-ODE}
  \frac{\dif{}^2}{\dif t^2}\c_\SO(t) \cdot m_{\SO} = \f_{s} 
\end{equation} 
where $m_{\SO}$ is the mass of the solid and $\f_s$ are the forces acting on
the solid in the horizontal and vertical directions. For simplicity, we assume
that the horizontal forces are negligible. The vertical forces are then the
gravitational pull, buoyancy and the viscous drag
\begin{equation}\label{eqn:Ball-Forces}
  \f_s =  \begin{pmatrix} 
            0\\0\\ m_{\SO}g - \vol(\SO)\rho_f g + \bm{F}_3,
          \end{pmatrix}
\end{equation}
where $\vol(\SO)$ is the volume of the solid and $\bm{F}_3$ the viscous drag
force in the vertical direction. This is the third component of
\begin{equation}\label{eqn:Drag-Lift}
  \bm{F} = \int_{\IN}\stress\n\dif s.
\end{equation}
Note that we added the effects of buoyancy in \eqref{eqn:Ball-Forces}. This
would be naturally included in $\bm{F}$ if we added body force $\rho_f g
\nabla \x_d$ to the right-hand side of the fluid equation
\eqref{eqn:Navier-Stokes-MomentumBallance}. However since this would only
affect the pressure, it is sufficient to consider the homogeneous equation
\eqref{eqn:Navier-Stokes-MomentumBallance} and add the effect of buoyancy to
\eqref{eqn:Ball-Forces}. Furthermore, this approach is more accurate in our
case, since only pressure-robust methods are able to reflect this effect
of gradient contributions in the forcing term on the pressure exactly on the
numerical level~\cite{JLM+17}.

As a result of us neglecting horizontal movement of the solid,
\eqref{eqn:Ball-ODE} becomes a scalar ODE. We can also simplify the terms in
\eqref{eqn:Ball-Forces}, so that in total we come to the equation 
\begin{equation}\label{eqn:solid-motion-ODE}
  \frac{\dif}{\dif t}\v_{\SO,3}(t) = \frac{\rho_s-\rho_f}{\rho_s}g +
    \frac{\bm{F}_3}{\vol(\SO)\rho_s}
\end{equation}
where $\v_\SO(t) = \frac{\dif}{\dif t}\c_\SO(t)$ is the solid's velocity.

The solid's motion couples back to the fluid equations through the boundary
condition at the interface $\IN$, by requiring continuity
of the velocity, i.e.,
\begin{equation}\label{eqn:continuity-vel-interface}
  \restr{\u}{\IN} = \v_{\SO} = \frac{\dif}{\dif t}\c_\SO(t).
\end{equation}
We note that this model neglects rotational effects. As will be shown below,
this does not have a major impact on the quality of the resulting
approximations.

\subsection{Domain Description}\label{subsec:domain}

\begin{figure}
  \hfil%
  \subfloat[][Three dimensions]{
    \includegraphics[scale=0.45]{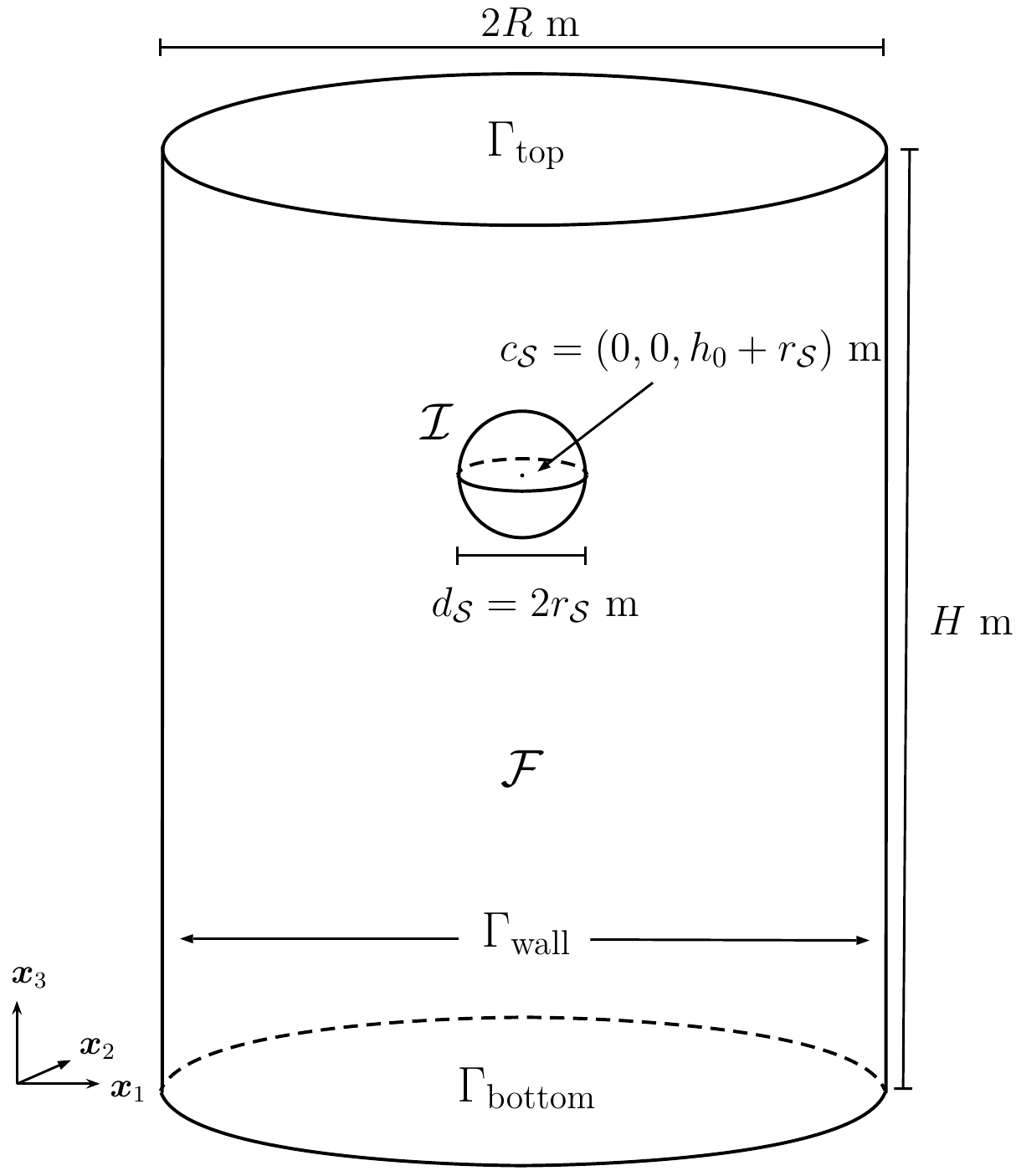}
    \label{fig:3d-initial-config}
  }%
  \hfil%
  \subfloat[][Roationally reduced]{
    \includegraphics[scale=0.45]{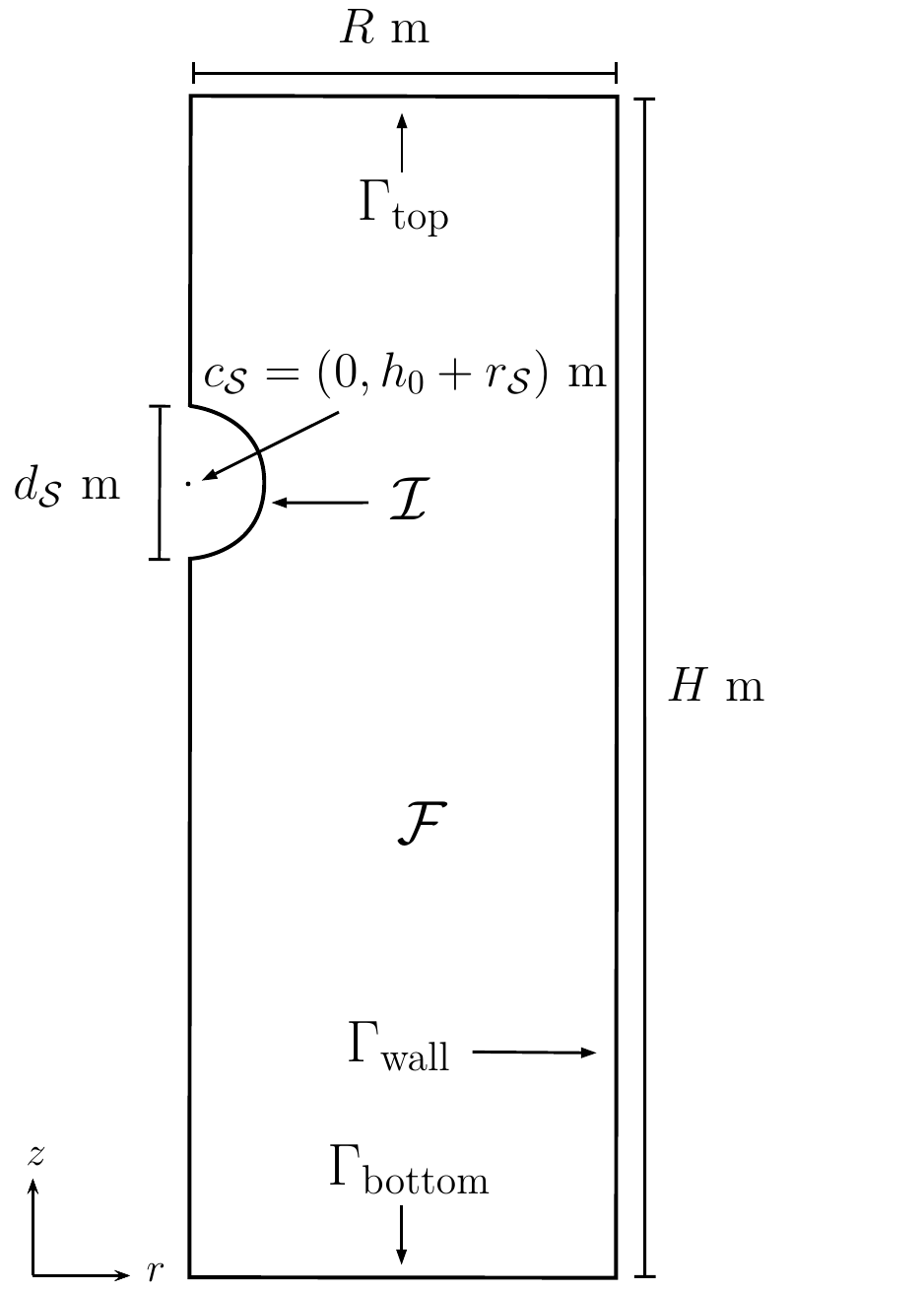}
    \label{fig:2.5d-initial-config}
  }%
  \hfil%
  \caption{The initial spatial configuration.}
  \label{fig:initial-config}
\end{figure}

Since we consider balls of different sizes, we shall keep the domain 
description general. The background domain is a cylinder $\O = \{\x=(x,y,z)^T
\in\R^3\st x^2 + y^2 < R^2, 0 < z < H\}$ for a given radius $R$ and a 
given height $H$. At $t=0$ the solid domain is described by $\SO(0) = \{\x=
(x,y,z)^T\in\R^3 \st x^2 + y ^2 + (z - (h_0 + r_\SO))^2 < r_\SO^2 \}$ for a
given ball radius $r_\SO$ and an initial height of the bottom of the ball
$h_0$.
Accordingly, the volume of the solid is given by $\vol(\SO) = 4\pi r^3 / 3$.
An illustration of this can be seen in
\hyperref[fig:3d-initial-config]{Figure~\ref*{fig:3d-initial-config}}.

\subsubsection{Boundary conditions}\label{subsubsec:bc}
We denote the top boundary ($z=H$) of the cylinder as $\Gtop$, the
bottom boundary ($z=0$) as $\Gbot$ and the side of the cylinder 
$(x^2 + y^2 = R^2)$ as $\Gwall$.

On the interface $\IN$ between the solid and the fluid, the Dirichlet boundary
condition is given by the continuity of the velocity, see
\eqref{eq:FSIcoupling} and
\eqref{eqn:continuity-vel-interface}. On the wall and bottom boundaries
$\Gbot\cup\Gwall$ we shall impose homogeneous Dirichlet boundary 
conditions $\u=\bm{0}$. In order to approximate the free surface at the top of
the water tank $\O$, we impose a free-slip boundary condition $\u_3=0$ at 
$\Gtop$. 

\subsection{Material parameters}

The geometrical and material parameters are as follows. The cylindrical tank
has a diameter of $110~\unit{mm}$ giving a radius of $R=0.055$\si{\metre} and 
has a height of $H=0.2$\si{\metre}.
The density, viscosity and surface tension of the highly viscous fluid are
$\rho_F = 1141~\si{\kilogram\per\metre\cubed}$,
$\mu_F=0.008~\si{\kilogram\per\metre\per\second}$ and
$\gamma_F=0.064~\si{\newton\per\metre\squared}$, respectively.

\subsubsection{PTFE6}\label{subsec:ptfe6}

We consider a polytetrafluorethylen/teflon (PFTE) particle with diameter
$d_\SO= \SI{6}{\milli\metre}$, i.e., $r_\SO=0.003$\si{\metre}., and a weight 
of $0.24~\si{\gram}$, which yields a particle density of $\rho_\text{PTFE} = 
2122~\si{\kilogram\per\metre\cubed}$. For the case of an  unconfined free 
fall, we calculated a terminal velocity of $v_T=0.38~\si{\metre\per\second}$ 
and a relaxation time of $t_R=0.1~\si{\second}$ according to~\cite{MMH15}. The
corresponding terminal Reynolds number is $Re_\text{PTFE}=333$.

Young's modulus and Poisson ratio of the PTFE ball are about 
$670 \si{\mega\pascal}=\SI{670000000}{\kilogram\per\metre\per\second\squared}$
\footnote{
\href{https://www.kugelpompel.at/upload/2312783_Datenblatt\%20Kunststoffkugel%
\%20PTFE\%20V1.01.pdf}{\texttt{www.kugelpompel.at/upload/2312783%
\_Datenblatt\%20Kunststoffkugel\%20PTFE\%20V1.01.pdf}}}
and $\nu_s=0.46$\footnote{\href{http://www.matweb.com/search/datasheet_print%
.aspx?matguid=4e0b2e88eeba4aaeb18e8820f1444cdb}{\texttt{www.matweb.com/search%
/datasheet\_print.aspx?matguid=4e0b2e88eeba4aaeb18e8820f1444cdb}}},
respectively, which yields the Lam\'e parameters $\lambda_s \approx
\SI[round-precision=5]{2638698630}{\kilogram\per\metre\per\second\squared},
$ and $\mu_s \approx
\SI[round-precision=5]{229452055}{\kilogram\per\metre\per\second\squared}$.
The spatial parameters are summarised in \autoref{tab:gemoetry} while the used
material parameters are summarised in \autoref{tab:physical-paramaters}.

\subsubsection{Rubber22}\label{subsec:rubber}

We consider a rubber sphere. This has a diameter of $d_\SO =
22~\si{\milli\metre}$ and a weight of $7.59~\si{\gram}$, combining for an
effective density of $\rho_\text{Rubber}= 1361~\si{\kilogram\per\metre\cubed}$.
The unconfined free fall characteristics
are $v_T=0.35~\si{\metre\per\second}$, $t_R=0.31~\si{\second}$ and 
$Re_\text{Rubber}=1109$.

Chemical analysis of the material suggests this to be hydrogenated nitrile
rubber (HNBR).
Young's modulus and Poisson ratio of this are approximately
$\SI[round-mode=off]{1.7}{} - \SI[round-mode=off]{20.7}{\mega\pascal} =
\SI{1700000}{}-\SI{20700000}{\kg\per\m\s\squared}$ \footnote{%
\href{https://eriks.de/content/dam/de/pdf/downloads/dichtungen/o-ringe/ERIKS
_Technisches-Handbuch-O-Ringe_de.pdf}{\texttt{eriks.de/content/dam/de/pdf%
/downloads/dichtungen/o-ringe/ERIKS\_Technisches-Handbuch-O-Ringe\_de.pdf}} 
p.\ 20.} %
and $\nu_s = 0.4999$~\cite{AMR04}. With $E_s=2\cdot 10^6$, we have the Lamé
parameters
$\lambda_s=\SI[round-precision=5]{3.33289e9}{\kg\per\m\per\s\squared}$ and
$\mu_s=\SI[round-precision=5]{6.66711e5}{\kg\per\m\per\s\squared}$.
These parameters are again summarised in \autoref{tab:gemoetry} and
\autoref{tab:physical-paramaters}.

\begin{table} 
  \centering
  \begin{tabular}{l cccl ccc}
    \toprule
    Experiment & \multicolumn{4}{c}{Geometry} & \multicolumn{3}{c}{Boundary
      Conditions}\\
    \cmidrule(r){2-5} \cmidrule(l){6-8}
      & R (\si{\metre}) & H (\si{\metre}) & $r_\SO$ (\si{\metre}) 
      & \multicolumn{1}{c}{$h_0$ (\si{\metre})} &
      $\Gwall \cup\Gbot$ & $\Gtop$ & $\IN$\\
    \midrule
    PTFE6 & \multirow{2}{*}{0.055} & \multirow{2}{*}{0.2} 
      & 0.003 & $0.1616616$
      & \multirow{2}{*}{$\u=0$} & \multirow{2}{*}{$\u_d=0$} 
      & \multirow{2}{*}{$\u=\v_\SO$} \\
    Rubber22 &&& 0.011 & $0.1461203$\\
    \bottomrule
  \end{tabular}
  \caption{Geometrical parameters of the test cases.}
  \label{tab:gemoetry}
\end{table}

\begin{table} 
  \centering
  \begin{tabular}{l c ccccc}
    \toprule
    Experiment & & \multicolumn{5}{c}{Material parameters}\\
    \cmidrule(lr){2-2} \cmidrule(lr){3-7}
      & $g$ (\si{\metre\per\second\squared}) & 
      $\mu_f$ (\si{\kilogram\per\metre\per\second}) & 
      $\rho_f$ (\si{\kilogram\per\metre\cubed}) &
      $\rho_s$ (\si{\kilogram\per\metre\cubed}) &
      $\lambda_s$ (\si{\kilogram\per\metre\per\second\squared}) & 
      $\mu_s$ (\si{\kilogram\per\metre\per\second\squared})\\
    \midrule
    PTFE6 & \multirow{2}{*}{$- 9.807$} & \multirow{2}{*}{$0.008$} &
      \multirow{2}{*}{$1141$} & $2122$ & $\SI[round-precision=5]{2638698630}{}$
      & $\SI[round-precision=5]{229452055}{}$ \\
    Rubber22 &&&& $1361$ & $\SI[round-precision=5]{3.33289e9}{}$ &
      $\SI[round-precision=5]{6.66711e5}{}$ \\
    \bottomrule
  \end{tabular}
  \caption{Summary of the benchmark set-up in standard units.}
  \label{tab:physical-paramaters}
\end{table}

\subsection{Quantities of interest}\label{subsec:quantities-interest}

We will use the following quantities to compare our numerical results with each
other and with the experimental data.

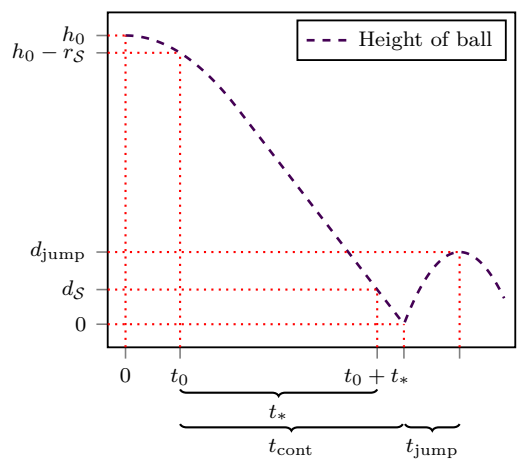
\begin{figure}
  \centering
  \input{plots/plot_definitions}
  \caption{Illustration for the definitions if the quantities of interest.}
  \label{fig:quantities-of-interest}
\end{figure}

\begin{description}[labelwidth=\widthof{$d_\text{jump}$}]
  \item[$t_\ast$] Let $t_0 = \restr{t}{\c_\SO = h_0}$  be the time at which the
    centre of mass is at $h_0$, i.e., when the ball has traveled $r_\SO$
    vertically. 
    We define $t_\ast$ as the time after release when $\dist(\IN, \Gbot) =
    d_\SO$ relative to $t_0$, i.e., for PTFE6 $t_\ast = \restr{t}{\c_\SO =
    (0,0,0.009)} - t_0$ and for Rubber22 $t_\ast = \restr{t}{\c_\SO =
    (0,0,0.033)} - t_0$.
  \item[$v_\ast$] The velocity of the ball in the $z$-direction at
    $t=t_\ast + t_0$.    
  \item[$f_\ast$] The vertical component of the force $\bm{F}$ acting on
    the ball at time $t_\ast + t_0$.
  \item[$t_\text{cont}$] The time of the first solid contact relative to $t_0$.
  \item[$t_\text{jump}$] The time between contact and the second change in 
    direction is realised, i.e., the amount of time the balls travels upwards 
    after the first contact.
  \item[$d_\text{jump}$] The maximum of $\dist(\IN, \Gbot)$ after contact,
    i.e., the size of the bounce.
\end{description}
An illustration of how these quantities are defined can be seen in
\autoref{fig:quantities-of-interest}.

\begin{remark}[Choice of reference values.]
In the experiments, we observed that the balls do not immediately start to fall
after they are released. 
The settling process starts with some distance to the liquid surface, but in
closest vicinity of the vacuum cup. In general, particles experience an
increased drag force when they are moving towards or away from a solid wall and
free fluid-surfaces. Accordingly, the early stage of the settling process here
is dominated by an increased drag force coming from the vacuum cup and the
liquid surface.

Due to the rather slow motion of the sphere at the beginning, the moment of
release cannot be defined well. To be able to compare the numerical
results with the experimental data, we therefore defined the above quantities
relative to the point in time at which the the ball has traveled the
distance of the balls radius.
Furthermore, since we have no measurement of the drag force in in the
experiment, $f_\ast$ will only be used to compare the computations directly.
\end{remark}

To establish $v_\ast, t_\text{cont}, d_\text{cont}$ and $d_\text{jump}$ from
the experimental data we interpolate the height data using a spline of order 3.
This spline is then evaluated to establish the time at which $\c_\SO = (0,0,
3r_\SO/2)$, the time of contact as well as the time and height of the jump. The
velocity is then taken as the first derivative of the spline representing the
height. The resulting reference values for the quantities of interest are
summarised in \autoref{tab:ref:experiment}.
  
\begin{table}
  \centering
  \begin{tabular}{l ccccc}
    \toprule
    Experiment & $t_\ast^\text{exp}$ & $v_\ast^\text{exp}$ &
      $t_\text{cont}^\text{exp}$ &
      $t_\text{jump}^\text{exp}$ & $d_\text{jump}^\text{exp}$ \\
    \midrule
    PTFE6 & $0.516403$ & $-0.330987$ & $0.492232$ & $0.027918$ & $0.002212$\\
    Rubber22 & $0.469137$ & $-0.309301$ & $0.544021$ & $0.089492$ &
      $0.004149$\\
    \bottomrule
  \end{tabular}
  \caption{Quantities of interest as established from the experimental data.}
  \label{tab:ref:experiment}
\end{table}

The experimental study~\cite{HTR20} was conducted, so that the horizontal
displacement of the particles was minimal. The data set we will compare
against shows a maximal horizontal displacement of less than
$\SI{2}{\milli\metre}$ and $\SI{0.75}{\milli\metre}$ in the PTFE6 and Rubber22
cases respectively.
This compares with the mean over time of the maximal deviation in the centre
location between experiment repetitions of $\SI{0.192}{\milli\m}$ and
$\SI{0.135}{\milli\m}$ for the PTFE6 and Rubber22 cases respectively.
However, the experiment is only able to give the projection of the horizontal
displacement onto the $x$-$z$-plane. Thus is is not possible to detect the true
horizontal motion.
Note that we have any horizontal motion this in the computation of the
reference values.
However, since overall the horizontal deflection is small, we consider this to
be reasonable for the present purpose.

%% file: plots/plot_definitions.tex
\begin{tikzpicture}
  \begin{axis}[
    width=7cm,
    ytick = {1, 0.94, 0.25, 0.12, 0},
    yticklabels = {$h_0$, $h_0 - r_\SO$, $d_\text{jump}$,$d_\SO$, 0},
    xtick = {0, 0.24495, 1.13, 1.25, 1.5},
    xticklabels = {0,$t_0$, $t_0+t_\ast$,,},
    tick align=outside,
    xmin = -0.08, xmax = 1.75,
    ymin = -0.08, ymax = 1.08,
    clip=false,
    ]    
    \addplot +[dashed] table [x=time, y=height, col sep=space]
        {plots/generic_fall_data.txt};
    \addlegendentry{Height of ball}
    
    \draw [thick, dotted, draw=red] 
      (axis cs: -0.08,1) -- (axis cs: 0,1);
    \draw [thick, dotted, draw=red] 
      (axis cs: 0,-0.08) -- (axis cs: 0,1);
    
    \draw [thick, dotted, draw=red] 
      (axis cs: -0.08,0.94) -- (axis cs: 0.24495,0.94);
    \draw [thick, dotted, draw=red] 
      (axis cs: 0.24495,-0.08) -- (axis cs: 0.24495,0.94);
            
    \draw [thick, dotted, draw=red] 
      (axis cs: -0.08,0.12) -- (axis cs: 1.13,0.12);
    \draw [thick, dotted, draw=red] 
      (axis cs: 1.13,-0.08) -- (axis cs: 1.13,0.12);
    
    \draw [thick, decoration = {brace, mirror, raise=5pt}, decorate] 
      (axis cs:0.24495,-0.18) -- 
        node[below=7pt] {\footnotesize $t_\ast$}
      (axis cs:1.13,-0.18);
    
    \draw [thick, dotted, draw=red] 
      (axis cs: -0.08,0) -- (axis cs: 1.25,0);
    \draw [thick, dotted, draw=red] 
      (axis cs: 1.25,-0.08) -- (axis cs: 1.25,0);
    \draw [thick, decoration = {brace, mirror, raise=5pt}, decorate] 
      (axis cs:0.24495,-0.3) -- 
        node[below=7pt] {\footnotesize $t_\text{cont}$}
      (axis cs:1.245,-0.3);
            
    \draw [thick, dotted, draw=red] 
      (axis cs: 1.5,-0.08) -- (axis cs: 1.5,0.25);
    \draw [thick, dotted, draw=red] 
      (axis cs: -0.08,0.25) -- (axis cs: 1.5,0.25);
    \draw [thick, decoration = {brace, mirror, raise=5pt}, decorate] 
      (axis cs:1.255,-0.3) -- node[below=7pt] {\footnotesize $t_\text{jump}$}
      (axis cs:1.5,-0.3);
  \end{axis}
  
\end{tikzpicture}

%% file: files/reduced_model.tex
\section{Reduced model}\label{sec:reduced-model}

The set-up described in \autoref{subsec:domain} is
symmetric with respect to rotation, if viewed in cylindrical
coordinates.
The experimental data presented in~\cite{HTR20,Hag20} shows a rotational
component in the motion of the solid and it also shows a small
deflection of the center of mass $c_\SO$ from the $z$-axis. 
However, these effects are small and since  the material parameters are such
that the resulting flow is the low to intermediate Reynolds-number
regime~\cite{HTR20}, we assume that the solution is described sufficiently well
by a rotationally symmetric flow in  cylinder coordinates.
We use this in order to obtain a two-dimensional reduced formulation which is
computationally cheaper compared to full three dimensional
computations. In \autoref{sec.ale.3d} we will also present a fully
resolved three dimensional simulation to have a closer look at these 
rotational effects. To distinguish between the full three dimensional domains
and the reduced two dimensional domain, we shall denote objects stemming from
the three-dimensional set-up with a superscript ``$3d$'' if there is a
potential for ambiguity. For the sake of readability, object based on the
reduced two-dimensional set-up will not have a special notation.

With the spaces $\V^{3d} = \{\v\in\H{1}{}{\FL^{3d}} \st \restr{\v}{\IN^{3d}} =
\u_{\SO^{3d}}\text{, } \restr{\v}{\Gwall^{3d} \cup \Gbot^{3d}} = 0 \text{ and }
\restr{\v_z}{\Gtop^{3d}} = 0\}\subset [ \pazocal{H}^1(\FL^{3d}) ]^3$  and
$Q^{3d}=\Ltwozero{\FL^{3d}}$, the weak formulation of the Navier-Stokes
equation
\eqref{eqn:Navier-Stokes} is as follows:
\begin{equation}\label{eqn:Navier-Stokes-3d-weak}
  \begin{gathered}
    \text{Find }(\u,p)\in \V^{3d}\times Q^{3d} \text{ such that for all } 
    (\v,q) \in \V^{3d}\times Q^{3d} \text{ it holds }\\
    A_f^{3d}(\u,p; \v,q) \coloneqq m^{3d}(\partial_t \u,\v) + a^{3d}(\u,\v) 
      + c^{3d}(\u,\u,\v) + b^{3d}(\v, p) + b^{3d}(\u, q) = 0
  \end{gathered}
\end{equation}
with the multilinear forms
\begin{equation}\label{eqn:forms}
  \begin{aligned}
    m^{3d}(\u,\v) &= \rho_f \int_{\FL^{3d}} \u\cdot\v\dif\x &
    a^{3d}(\u,\v) &= \mu_f\int_{\FL^{3d}} \nabla\u : \nabla \v \dif\x \\
    c^{3d}(\u,\v,\w) &= \rho_f \int_{\FL^{3d}} (\u\cdot\nabla)\v\cdot\w\dif\x &
    b^{3d}(q,\v) &= -\int_{\FL^{3d}} q \nabla\cdot\v\dif\x.
  \end{aligned}
\end{equation}

In order to reduce this three dimensional flow problem into a two dimensional
flow problem, we rewrite the problem into cylinder coordinates $r,
\phi, z$. Rotational symmetry of a solution $\u$ means that
$\partial_\phi\u=0$. We use this to rotate the domain into the $r^+-z$-plane
and transform the weak formulation \eqref{eqn:Navier-Stokes-3d-weak}
accordingly. Now let $\nabla = \left( \begin{smallmatrix}
\partial_r\\ \partial_z \end{smallmatrix}\right)$ and $\O =
\O^{3d}\cap(\R^+\times\R)$ be the reduced two dimensional computational domain.
A sketch of the three dimensional domain transformed into two dimensions can be
seen in
\hyperref[fig:2.5d-initial-config]{Figure~\ref*{fig:2.5d-initial-config}}.

At the symmetry boundary ($r=0$) the boundary conditions 
\begin{align}\label{eq:symbc}
  \u_r =0, \qquad \partial_r \u_z =0
\end{align}
are valid. The reduced spaces are then $\V = \{ \v\in\H{1}{}{\FL} \st
\restr{\v}{\IN} =  \left(\begin{smallmatrix} 0 \\ {\v_\SO}_3
\end{smallmatrix}\right)\text{, } \restr{\v}{\Gwall\cup\Gbot} = 0 \text{, }
\restr{\v_z}{\Gtop} = 0 \text { and } \restr{\v_r}{r=0} = 0
\}\subset [ \pazocal{H}^1(\FL)]^2$ and $ Q = \Ltwozero{\FL}$.
The reduced Navier-Stokes problem then reads
\begin{equation}\label{eqn:Navier-Stokes-2.5d-weak}
  \begin{gathered}
    \text{Find }(\u,p)\in \V\times Q \text{ such that for all } 
    (\v,q) \in \V \times Q \text{ it holds }\\
    A_f(\u,p; \v,q) \coloneqq  m(\partial_t \u,\v) + a(\u,\v) + c(\u,\u,\v)
      + b(\v, p) + b(\u, q) = 0
  \end{gathered}
\end{equation}
with the transformed multilinear forms
\begin{align*}
  m(\u,\v) &= 2\pi\rho_f \int_{\FL}r \u\cdot\v\dif\x &
  a(\u,\v) &= 2\pi\mu_f\int_{\FL}\nabla\u:\nabla\v+\frac{1}{r}\u_r\v_r\dif\x\\
  c(\u,\v,\w) &= 2\pi \rho_f \int_{\FL}r(\u\cdot\nabla) \v\cdot \w\dif\x &
  b(q,\v) &= -\int_{\FL} q (\v_r + r\nabla\cdot\v)\dif\x
\end{align*}
c.f.~\cite{BDM99}. Since the motion of the solid is partially driven by the
fluid forces acting on it, we also need to transform these. Let 
$(\u^{3d}, p^{3d})\in\V^{3d}\times Q^{3d}$ be a rotationally symmetric solution
of \eqref{eqn:Navier-Stokes-3d-weak} and $(\u, p)\in\V\times Q$ the solution
of \eqref{eqn:Navier-Stokes-2.5d-weak}. We can then transform the weak
boundary integral formulations as
\begin{equation}\label{eqn:rotated-stress}
  \int_{\IN^{3d}} \bm{\sigma}^{3d}(\u^{3d},p^{3d})\n^{3d}\cdot\v^{3d}
    \dif S^{3d}
  = 2\pi \int_{\IN}r(\mu_f \nabla\u - \Id p)\n\cdot\v\dif S.
\end{equation}
The forces $\bm{F}$ can then be computed by inserting the appropriate
non-conforming test-functions $\v$ into this functional. Using this,
we can compute the motion of the solid as before using
\eqref{eqn:solid-motion-ODE}.

%% file: files/numerical_experiment.tex
\section{Numerical computations}\label{sec:num-comp}

We give the details of the different numerical approaches applied to attempt to
reproduce the observed data and present the results attained with these
methods.

The full result data-sets of all methods and the source code to reproduce the
rigid body ALE and CutFEM methods can be found in the zenodo repository
\textsc{doi}:~%
\href{https://doi.org/10.5281/zenodo.3989604}{\texttt{10.5281/zenodo.3989604}}.
This also includes the preparatory examples presented in \autoref{appendix}.

\subsection{Discretisations}\label{subsec:discretisations}

We provide the details on the formulations of the different discretisation
approaches used.

\input{files/ale}

\input{files/cutfem}

\input{files/fsi}

\input{files/results}

%% file: files/ale.tex
\subsubsection{Fluid-Rigid Body interaction in Arbitrary Lagrangian-Eulerian
coordinates}
\label{sec.ale}

To compute the spatially reduced coupled problem we formulate the Navier-Stokes
equations in Arbitrary Lagrangian Eulerian (ALE)
coordinates~\cite{HughesLiuZimmermann1981} by introducing a reference map
\[
T_{ALE}(\c_\SO):\FL\to\FL(\c_\SO),
\]
where $\c_\SO$ is solid's centre of mass relative to the bottom boundary
and where
\[
\FL(\c_\SO) = \big\{[0,\unit[0.055]{m}]\times[0,\unit[0.2]{m}]\big\}
\setminus B_{r_\SO}\big(\c_\SO\big),
\]
$B_{r_\SO}(\c_\SO)$ being the open ball of radius $r$ with midpoint $\c_S$. As
reference domain we set $\FL \coloneqq \FL(\unit[0.05]{m})$ where the
ball is centred in $\c_\SO=(0,0,0.5)$ such that the mesh distortion is
not too extreme when the solid comes close to the lower boundary. The mapping
$T_{ALE}(\c_\SO)$ is given analytically by
\begin{equation}\label{ALEmap}
  \begin{aligned}
    T_{ALE}(\c_\SO;r,z)& =
    \Big(r,z+(\c_\SO-\unit[0.05]{m})f_{ALE}(z)\Big),\\
    f_{ALE}(z) &= \begin{cases}
      \frac{z}{\unit[0.05]{m}-2r_\SO}& z<\unit[0.05]{m}-2r_\SO\\
      1 & \unit[0.05]{m}-2r_\SO \le z\le \unit[0.05]{m}+2r_\SO\\
      \frac{\unit[0.2]{m}-z}{\unit[0.15]{m}+2r_\SO}&z>\unit[0.05]{m}+2r_\SO\\
    \end{cases}
  \end{aligned}
\end{equation}
such that $T_{ALE}(\c_\SO;r,z)$ is a pure translation for all $z\in
[\unit[0.05]{m}-2r_\SO,\unit[0.05]{m}+2r_\SO]$, where the reference ball
is located. We make sure that the lines at $z=\unit[0.05]{m}\pm 2
r_\SO$ are resolved by the computational mesh such that $T_{ALE}$ is
differentiable within all elements. This construction of the domain
map does not allow us to reduce the distance of the ball from the
lower boundary to less than $r_\SO$. 

The reference domain $\FL$ is basis for the finite element
discretisation.  In cylindrical coordinates, the ALE version of the
variational formulation takes the form
\begin{align*}
  m_{ALE}(\c_\SO;\u,\v) &= 2\pi\rho_f \int_{\FL}rJ(\c_\SO)
    \u\cdot\v\dif\x \\
  a_{ALE}(\c_\SO;\u,\v) &= 2\pi\mu_f\int_{\FL}J(\c_\SO)
    \nabla\u\begin{pmatrix} 1 &0\\ 0&J(\c_\SO)^{-1}\end{pmatrix}:
    \nabla\v\begin{pmatrix} 1 &0\\ 0&J(\c_\SO)^{-1}\end{pmatrix}
    + \frac{J(\c_\SO)}{r}\u_r\v_r\dif\x \\
  c_{ALE}(\c_\SO;\u,\v,\w) &= 2\pi \rho_f \int_{\FL} rJ(\c_\SO) \nabla
    \begin{pmatrix} 1 &0\\ 0&J(\c_\SO)^{-1} \end{pmatrix}
    \big(\u-\partial_t T_{ALE}(\c_\SO)\big)\cdot \w\dif\x \\
  b_{ALE}(\c_\SO;q,\v) &= -2\pi\int_{\FL} J(\c_\SO) q \big(\v_r + r
    \partial_r\v_1+\partial_z\v_2/J(\c_\SO)\big) \dif\x,
\end{align*}
where $J(\c_\SO) = \operatorname{det}(\nabla T_{ALE})$ is the determinant
of the deformation gradient.

The discretisation is realised by means of quadratic equal-order finite
elements for
velocity and pressure on a quadrilateral mesh of the reference domain $\FL$, we
refer to~\cite[Section~4.4]{Richter2017} for details on the realisation. To
stabilise the inf-sup condition we use the local projection scheme as
introduced by Becker and Braack~\cite{BeckerBraack2001}, given, in ALE
formulation, as
\begin{align*}
  s_{ALE}(\c_\SO;p,q) &=
  \int_{\FL}r
  \begin{pmatrix}
    \delta_r &0\\0&\delta_z
  \end{pmatrix}\nabla \kappa_h (p)\cdot
  \begin{pmatrix}
    J(\c_\SO)&0\\0&J(\c_\SO)^{-1}
  \end{pmatrix}\nabla \kappa_h(q)
  +\frac{J(\c_\SO)}{r}\delta_r \kappa_h(p)\kappa_h(q)
  \dif\x.
\end{align*}
Here $\kappa_h\coloneqq \operatorname{id}-i_{2h}$ is the coarse mesh
fluctuation operator that subtracts the interpolation to the mesh with
double spacing and where $\delta_r$ and $\delta_z$ are local stabilisation
parameters that depend on the element diameter $h$ and the time step size $k$
\[
\delta_r = 0.1\cdot\left(\frac{\mu_f}{\rho_f h^2}+\frac{1}{k}\right)^{-1},\quad
\delta_z = 0.1\cdot\left(\frac{\mu_f}{\rho_f h^2
J(\c_\SO)^2}+\frac{1}{k}\right)^{-1}.
\]
The different scaling in $r$- and $z$-direction reflects the anisotropy induced
by the ALE transformation, c.f.~\cite{BraackRichterEnumath2006}
or~\cite[Section~5.3.3]{Richter2017}.

In time we discretise the Navier-Stokes equations and the rigid body problem
with BDF2 time stepping in a decoupled iteration taking $10^{-8}$ as tolerance
for the solid velocity and deformation update. The forces acting on
the solid are evaluated by means of the Babu\vs{s}ka-Miller
Trick~\cite{BabuskaMiller1984,vWRL+19} such that quadratic finite
elements let us expect fourth order convergence. Hence, each step of
spatial refinement will be accompanied by two refinement of the time step.

\paragraph{Fluid-Rigid Body interaction in Arbitrary Lagrangian-Eulerian
coordinates in three dimensions}

To clarify the role of fixed body rotations of the solid and
deflections of the center of mass from the $z$-axis, both of which were
observed in the experimental analysis~\cite{HTR20,Hag20}, we perform
full three dimensional simulations. The Navier-Stokes equations are
formulated in Cartesian coordinates~\eqref{eqn:Navier-Stokes} and the
motion of the rigid body is described by 
\begin{equation}\label{Eqn:body:3d}
  \begin{aligned}
    \frac{\dif{}}{\dif t}\c_\SO(t)  = \v_\SO(t),\quad
    \frac{\dif{}}{\dif t}\v_\SO(t) \cdot m_{\SO}
    &= \begin{pmatrix}
      0 \\ 0 \\ m_\SO g -\vol(\SO)\rho_fg 
    \end{pmatrix}
    +\int_{\IN}\stress\n\dif s,\\
    \frac{\dif{}}{\dif t}\omegat_\SO(t) \cdot I_{\SO}
    +\omegat_\SO(t)\times I_\SO\omegat_{\SO}(t)
    &=
    \int_{\IN}\big({\bm x}-\c_\SO(t)\big)\times \stress\n\dif s,\\
  \end{aligned}
\end{equation}
where $\v_\SO(t)\in\mathds{R}^3$ is the full velocity vector,
$\c_\SO(t)\in\mathds{R}^3$ the solid's center of mass and
$\omegat_\SO(t)\in\mathds{R}^3$ the rotational velocity vector,
c.f.\ \eqref{eqn:Ball-ODE}, \eqref{eqn:Ball-Forces},
\eqref{eqn:Drag-Lift}. 
Assuming a homogenous distribution of the density the moment of inertia is
given by
\begin{equation*}
  I_\SO = \frac{8}{15}\pi \rho_\SO r_\SO^5 \Id.
\end{equation*}
It follows that the nonlinear term in the rotational ODE \eqref{Eqn:body:3d}
vanishes, since
$\omegat_\SO(t)\times I_\SO\omegat_{\SO}(t)
= \frac{8}{15}\pi \rho_\SO r_\SO^5~\omegat_\SO(t)\times \omegat_{\SO}(t)
= 0$. 
On the surface of the sphere, the Navier-Stokes Dirichlet conditions
are used to describe the velocity
\[
\v({\bm x},t)= \v_\SO(t) + \omegat_\SO(t)\times \big({\bm x}-\c_\SO\big)
\text{ on } \IN. 
\]
To evaluate the torque correctly through the variational formulation, i.e., by
using the Babu\vs{s}ka-Miller Trick~\cite{BabuskaMiller1984,vWRL+19},
the symmetric gradient is used instead in~\eqref{eqn:forms}, hence
\[
a^{3d}(\u,\v) = \mu_f\int_{\O^{3d}} (\nabla\u +\nabla\u^T) : \nabla
\v \dif\x. 
\]
We cast the problem in standard ALE formulation and refer
to~\cite[Chapter~5]{Richter2017} for further details. The ALE map $T_{ALE}(t)$
is chosen similar to the reduced case~\eqref{ALEmap}, but we must
incorporate motion in the $x$-$y$-plane and define
\[
T_{ALE}(\c_\SO;x,y,z) =
\begin{pmatrix}
  x + \c_{\SO,1}\cdot g_{ALE}\big(r(x,y)\big)\\
  y + \c_{\SO,2}\cdot  g_{ALE}\big(r(x,y)\big)\\
  z+(\c_{\SO,3} - r_\SO - \unit[0.05]{m})f_{ALE}(z),
\end{pmatrix}
\]
where $f_{ALE}(\cdot)$ is defined in \eqref{ALEmap} while
$r(\cdot,\cdot)$ and $g_{ALE}(\cdot)$ are given by
\[
r(x,y) = \max\left\{0,
\min\left\{1,\frac{x-r_\SO}{R-r_\SO}\right\}\right\},\quad
g_{ALE}(r) = 1-\frac{1}{1+\exp\left(\frac{1-2r}{r-r^2}\right)}.
\]
The function $r(\cdot,\cdot)$ maps the $x/y$ plane to $[0,1]$ with
$r(x,y)=0$ for $\sqrt{x^2+y^2}\le r_\SO$ and $r(x,y)=1$ for
$\sqrt{x^2+y^2}\ge R$. Further, $g_{ALE}$ is a smooth transition
function mapping $(0,1)$ to $(0,1)$ with all derivatives being zero at
$0$ and $1$. On the other hand, the derivative of
$f_{ALE}(z)$ is not defined at $z=\unit[0.05]{m}\pm 2r_\SO$. These two
surfaces are however resolved by the finite element mesh to give good
approximation characteristics of the ALE formulation.

\paragraph{Implementation}

Both formulations, the reduced two dimensional ALE formulation in
cylindrical coordinates and the full three dimensional ALE formulation
are implemented in the finite element library 
\texttt{Gascoigne 3D}~\cite{Gascoigne}.
The coupling between the Navier-Stokes equation and
the rigid body motion is resolved in a simple iteration until the
discrepancy in velocity reached a threshold
\[
\|\v\big|_\IN-\v_\SO \|_\infty < 10^{-8}. 
\]
The nonlinearity of the Navier-Stokes equation is solved by a Newton
iteration and the resulting linear systems of equations are
approximated with a parallel GMRES iteration, preconditioned by a
geometric multigrid solver, see~\cite{FailerRichter2020}. The meshes
are graded with a higher resolution close to the solid.

%% file: files/cutfem.tex
\subsubsection{Fluid-Rigid Body interaction in Eulerian coordinates}
\label{sec.cutfem}

As in \autoref{sec.ale}, we consider the problem as a moving domain problem
for the fluid, assume the solid to be a rigid body and decouple the
fluid and solid equations.
For the resulting moving domain problem we use an unfitted Eulerian finite
element method from~\cite{WRL20} using BDF2 time-stepping, together with
Taylor-Hood elements in space, which are inf-sup stable in the CutFEM
setting~\cite{GO17} with ghost-penalty
stabilisation~\cite{Bur10}. In order to obtain the full convergence order of
the Taylor-Hood finite element pair, we use an isoparametric mapping introduced
in~\cite{Leh16} for stationary domains to realise the higher-order geometry
approximation.

\paragraph{Transformed Nitsche terms}
In the used CutFEM method, boundary conditions on unfitted boundaries are
enforced using Nitsches method~\cite{Nit71}. These also need to be transformed
to the rotationally symmetric setting
such that we have a consistent and stable formulation.

Let $\G$ be the Dirichlet boundary. Integrating the diffusion operator
$a(\cdot, \cdot)$ by parts in cylinder coordinates, we find that the
consistency term is
\begin{equation*}
  n_c(\u,\v) = - 2\pi \mu_f\int_{\G} r(\nabla\u)\n \cdot \v \dif S.
\end{equation*}
To enforce the Dirichlet boundary conditions consistently, we also scale the
penalty term with $r$, such that we have
\begin{equation*}
  n_s(\u,\v) = 2\pi \mu_f \sigma \frac{k^2}{h} \int_{\G} r\u\cdot\v \dif S.
\end{equation*}
with a penalty parameter $\sigma > 0$, the velocity space's polynomial order
$k$ and the local mesh diameter $h$. In total we have the consistent and
symmetric Nitsche formulation of the vector valued diffusion operator
$a_h(\u,\v) = a(\u,\v) + n_c(\u, \v) + n_c(\v, \u) + n_s(\u, \v)$.
Similarly we find for the pressure-coupling operator, that the Nitsche term is
\begin{equation*}
  n_p(\v,q) = \int_{\G} r q \v\cdot\n \dif S
\end{equation*}
so that the consistent reduced formulation is $b_h(\v, q) = b(\v, q) + 
n_p(\v,q)$. To implement inhomogeneous Dirichlet conditions, we then add the 
symmetry and penalty terms with the required boundary values to the right-hand
side of the fluid equations.

\paragraph{Transformed ghost-penalty operators}
The role of the ghost-penalty operator in the unfitted Eulerian time-stepping
scheme used here is twofold. As in other CutFEM discretisations, it stabilises 
arbitrary element cuts, such that the method is stable and the resulting
matrices are well-conditioned~\cite{Bur10}. 
Secondly, appropriately scaled ghosty-penalties provide the necessary implicit
extension for the method-of-lines approach to the discretisation of the 
time-derivative~\cite{LO19,BFM19,WRL20}. 

We use the \emph{direct-version} of the ghost-penalty operator~\cite{Pre18}. 
Let $\Fh$ be a set of facets between neighbouring elements on which the
ghost-penalty operator is to act on, c.f.~\cite{LO19, WRL20} for further
details.
For a facet $F\in\Fh$ such that $F=\overline T_1 \cap \overline T_2$, we define
the facet patch $\omega_F = T_1 \cup T_2$. The velocity ghost-penalty operator
for the rotationally symmetric formulation is then
\begin{equation*}
  i_h(\u,\v) = \gamma_{\u} \sum_{F\in\Fh}\int_{\omega_F} \frac{r}{h^2}(\u_1 -
    \u_2)\cdot (\v_1 - \v_2) + \frac{1}{r}(\u_{r,1} -\u_{r,2})(\v_{r,1}
    -\v_{r,2}) \dif\x
\end{equation*}
where $\v_i=\restr{\E\v}{T_i}$ with the canonical extension of polynomials
$\E:\P{}{T}\rightarrow\P{}{\R^d}$. The pressure ghost-penalty operator is
\begin{equation*}
  j_h(p,q) = \gamma_{p} \sum_{F\in\Fh}\int_{\omega_F} r(p_1 -p_2)(q_1 - q_2)
    \dif\x.
\end{equation*}
With these transformed ghost-penalty operators, it is easy to show the standard
ghost-penalty results in the appropriately transformed norms.

\paragraph{Contact algorithm}
We consider a very basic contact avoidance scheme, used widely in the
literature~\cite{GPHJ99, WT06, ST08, Fre16}. The idea is to introduce an
artificial (lubrication) force acting on the rigid body in the vicinity of the
contact wall, which increases the closer the ball comes to the wall and acts in
the direction away from the wall. This force is then added to the forces
governing the motion of the rigid solid such that contact does not occur. We
define this force as
\begin{equation*}
  f_c(\SO) = \begin{cases}
    0 & \mbox{if } \dist(\IN, \Gbot) \geq dist_0\\
    \gamma_c \frac{dist_0 - \dist(\IN, \Gbot)}{\dist(\IN,
      \Gbot)} & \mbox{if }\dist(\IN, \Gbot) < dist_0
   \end{cases}
\end{equation*}
where $dist_0$ and $\gamma_c$ are parameters to be chosen and $\dist(\IN,
\Gbot)$ is the minimal distance between $\IN$ and $\Gbot$. We then add this to
the right-hand side of the ODE \eqref{eqn:Ball-ODE} and carry this thought, so
that the right-hand side of \eqref{eqn:solid-motion-ODE} becomes
$\frac{\rho_\SO-\rho_\FL}{\rho_\SO}g + \frac{\bm{F}_y+f_c}{\vol(\SO)\rho_\SO}$.

\paragraph{Implementation}

This discretisation is implemented using the finite-element library
\texttt{netgen}/\texttt{NGSolve}, see~\cite{Sch97, Sch14} and
\href{https://ngsolve.org}{\texttt{ngsolve.org}}, together with the add-on
package \texttt{ngsxfem}~\cite{ngsxfem} for unfitted finite element
functionality.

The background mesh is constructed by defining a local mesh parameter
$h_\text{inner}$ on the left of the reduced domain where $r < 2 d_\SO/3$ and
then creating a shape regular mesh with $h=\hmax$ in the remainder of the
domain. This is to obtain more accurate boundary integrals, i.e., when
computing the forces acting on the ball. In the Rubber22 setting we choose
$h_\text{inner} = \hmax/4$ and in the PTFE6 case as $h_\text{inner}=\hmax/7$.
This is to obtain background meshes with a similar number of elements in each
settings. On the active part of the mesh we consider standard $\PP^2/\PP^1$
Taylor-Hood elements. 

The Nitsche parameter is taken as $\sigma=100$, the extension ghost-penalty
parameter is $\gamma_{\u, e} = 0.1$ and the cut-stability ghost-penalty
parameter $\gamma_{\u, s} = \gamma_{p,s} = 0.01$ and the extension strip
width parameter is $c_\delta=4$. See~\cite{WRL20} for details on these
parameters. 

The contact parameters are tuned with respect to the PTFE6 jump height, since
the model cannot be expected to resolve the elastic nature of rubber. For the
PTFE6 we take the contact model parameters $dist_0 = 2\cdot 10^{-5}$ and
$\gamma_c= 0.38$.
Since the mass of the Rubber22 ball as approximately $31.6$ times larger than
the PTFE6 ball, we take contact model parameters which are appropriately
larger, such that the resulting acceleration acting on the balls is comparable.
For the Rubber22 computations we take $dist_0=2\cdot 10^{-5}$ and $\gamma_c
= 12$.

Each time-step is iterated between the fluid system and the solid ODE until the
system is solved. We consider the system as solved, when the update of the
balls velocity in an iteration is less than $10^{-8}$.

%% file: files/fsi.tex
\subsubsection{Fluid-Structure interaction in fully Eulerian coordinates}
\label{sec.fsi}

We consider the full fluid-structure interaction problem including contact with
$\Gbot$.
We adopt here a fully Eulerian approach for the FSI system in order to enable
the transition to contact~\cite{Dunne2006, Richter2012b, Fre16}.
To allow for an implicit inclusion of the contact conditions into the system
(see below), we use a Nitsche-based method for the FSI coupling as presented
in~\cite{BurmanFernandezFrei2020}.

\paragraph{Solid bilinear form and FSI coupling}
The fluid bilinear form has already been detailed
in~\eqref{eqn:Navier-Stokes-2.5d-weak}.
To introduce the solid form, 
we denote the reduced solid domain by ${\SO}(t)$ and define the bilinear form
corresponding to the linear elasticity equations in Eulerian
coordinates~\eqref{eq:linElast} by
\begin{align*}
  A_s(\d,\dot{\d}; \w,\z) := {m}_s(\partial_t \dot{\d} - \dot{\d}\cdot {\nabla}
    \dot{\d}, \w) + {a}_s(\d,\w) + m_{\dot{\d}}(\partial_t \d -\dot{\d}\cdot
    \nabla \d + \dot{\d}, \z),
\end{align*}
where 
\begin{align*}
  {m}_s(\d, \w) := 2\pi\rho_s \int_{\SO}r \d\cdot\w\dif\x,
    &\qquad{a}_s(\d,\w) := 2\pi\int_{\SO} r \nabla \sigma_s(d) :{\nabla} \w 
    +{\sigma}_{s,r} \w_r \dif\x,\\
  m_{\dot{d}}(\d,\w) &:= 2\pi \int_{\SO}\d\cdot\w\dif\x.
\end{align*}
and 
\begin{align*}
  {\sigma}_s = 2 \mu_s E(\d) +\lambda_s \left( \text{tr}(E(\d)) + \frac{1}{r}
    \d_r\right) I,\qquad
  {\sigma}_{s,r} =\frac{2\mu_s+\lambda_s}{r} \d_r + \lambda_s \text{tr}(E(\d)).
\end{align*}

Moreover, we make use of the Nitsche terms defined in \autoref{sec.cutfem}
to impose the FSI coupling conditions~\eqref{eq:FSIcoupling}
\begin{align*}
  n(\u,p,\dot{\d}; \v, q,\w) := n_s(\u-\dot{\d}, \v-\w) + n_c(\u,\v-\w) +
    n_c(\v,\u-\dot{\d}) + n_p(\v-\w, p) - n_p(\u-\dot{d}, q).
\end{align*}
Note the negative sign in front of the last term, which is required to ensure
stability of the FSI formulation, see~\cite{BurmanFernandez2014}.

\paragraph{Discretisation and stabilisation}

In order to resolve the interface $\IN$ within the discretisation, we use the
locally modified finite element method introduced in~\cite{FreiRichter14}. This
\textit{fitted} finite element method is based on a coarse \textit{unfitted}
patch mesh, which is independent of the interface location. The coarse cells
are then divided in such a way into sub-triangles and sub-quadrilaterals that
the interface is resolved in a linear approximation. In this work, we use
equal-order locally modified finite elements of first order, in combination
with an anisotropic edge-oriented pressure stabilisation term $s_p(p,q)$,
see~\cite{Frei2019} for the details.
We denote the locally modified finite element space of first order by $X_h^1$.
The discrete spaces for fluid velocity and pressure $\V_h$ and $ Q_h$ and for
solid displacement and velocity $\W_h$ and $ Z_h$ are given by applying the
respective Dirichlet conditions to the locally modified finite element space
$X_h^1$ and by restricting the degrees of freedom to the fluid or solid
sub-domain, respectively.

In addition, we add a stabilisation term $s_{\d}(\dot{d},z)$ 
of artificial diffusion
type to the solid equations, see~\cite{Fre16}, 
and a consistent stabilisation at the boundary $\Gamma_{\text{sym}}$
corresponding to the
rotational axis to ensure that the second boundary condition~\eqref{eq:symbc}
is accurately imposed in the discrete formulation
\begin{align*} 
s_{\d}(\dot{d},\z) = \alpha_d h^2 (r {\nabla} \dot{d}, {\nabla}\z)_{\SO}, \qquad
  s_r(\u,\v) := \alpha_{\text{sym}} \rho_f\mu_f \int_{\Gamma_{\text{sym}}}
    h_n^2 h_\tau \partial_r \u_z \partial_r \text{d}{S}.
\end{align*}
Here, $h_n$ and $h_\tau$ refer to the cell-sizes in normal and tangential
direction, respectively.
We summarise the stabilisation terms in the bilinear form
\begin{align*}
  s(\u,p,\dot{d}; \v,q,\z) = s_p(p,q) + s_{\d}(\dot{d},\z) +
    s_r(\u,\v).
\end{align*} 
 
For time discretisation, we use the modified dG(0) time discretisation
presented in~\cite{FreiRichter2017}, which can be seen as a variant of the
dG(0)/backward Euler methods that considers the movement of the interface in
each space-time slab. The interface positions and the domain affiliations are
updated explicitly based on the displacement $d(t_{n-1})$ of the previous
time-step and using the Initial Point Set/Backwards Characteristics
method~\cite{Dunne2006, Fre16}. This means that we set
\begin{align*}
  \IN^n := \IN(d(t_{n-1})), \quad 
  \SO^n := \SO(d(t_{n-1})), \quad 
  \FL^n :=\FL(d(t_{n-1})).
\end{align*}

\paragraph{Contact treatment}

When a part $\IN_c:=\IN\cap \Gbot$ of $\IN$ enters into contact with $\Gbot$,
the FSI conditions need to be substituted with appropriate contact conditions.
It has been noted in~\cite{BurmanFernandezFrei2020} and~\cite{AgerWalletal}
that although the fluid layer between ball and lower wall vanishes (from a
macroscopical perspective), an extension of the fluid forces to the contact
surface $\IN_c$ has to be considered to obtain a physically relevant contact
formulation. Here we use the simplest possible numerical approach, which is to
relax the no-penetration condition by a small $\epsilon=\epsilon(h)>0$, such
that a very thin mesh-dependent fluid layer remains at all times.

The distance to $\Gbot$ depends on the (Eulerian)
displacement $\d(t_n)$ in the following way 
\begin{align*}
  \text{dist}\left(\IN(t_n), \Gbot\right) = \text{dist}\left(\IN(t_{n-1}),
    \Gbot\right) + \d_n(t_n) - \d_n(t_{n-1}).
\end{align*}
The contact conditions, relaxed by a small $\epsilon=\epsilon(h)>0$, read
\begin{equation}\label{ContactCond}
  \text{dist}\left(\IN(t_n), \Gbot\right)\leq \epsilon, \quad \jump{\sigma_n}
    \leq 0, \quad 
  \jump{\sigma_n} \left(\text{dist}\left(\IN(t_n), \Gbot\right)
    -\epsilon\right)  = 0 \quad 
  \text{ on } \IN^n = \IN(t_{n-1}),
\end{equation}
where
\begin{align*}
  \jump{\sigma_n}:= n^T {\sigma}_s n - n^T {\sigma}_f n.
\end{align*}
In other words the relaxation means that the contact conditions are already
applied at an $\epsilon$-distance from $\Gbot$.
The three conditions
\eqref{ContactCond} can be equivalently formulated in equality form with an
arbitrary $\gamma_C>0$~\cite{AlartCurnier91}
\begin{align}\label{ContactCond_eq}
  \jump{\sigma_n} = -\gamma_C \Big[  \underbrace{\text{dist}\left(\IN(t_n),
    \Gbot\right)- \epsilon -\gamma_C^{-1}
    \jump{\sigma_n}}_{=:P_\gamma(\d,\jump{\sigma_n})} \Big]_+ \quad \text{ on }
    \IN^n,
\end{align}
where $[\cdot]_+$ stands for the positive part of $(\cdot)$. The contact
parameter $\gamma_C$ will be chosen as $\gamma_C= \gamma_C^0 \lambda_s h^{-1}$,
the
relaxation parameter as $\epsilon=\epsilon_0 h_y$, where $h_y$ denotes the cell
size in vertical direction at the bottom of the cylinder.

Note that \eqref{ContactCond_eq} includes both the FSI coupling and the contact
condition, as in absence of contact, it is exactly the FSI interface condition
in normal direction. For this reason the transition between FSI coupling and
contact conditions can be included easily in a fully implicit fashion in the
variational formulation.

The final variational formulation reads:
\begin{equation*}
  \begin{gathered}
    \text{Find }(\u,p,\d,\dot{\d})\in \V_h\times  Q_h\times \W_h \times \Z_h
    \text{, such that for all } (\v,q,\w,\z) \in \V_h \times  Q_h\times \W_h
    \times \Z_h \text{ it holds}\\
    \begin{aligned}
      A_f(\u,p; \v,q) + A_s(\d,\dot{\d}; \w,\z) + n(\u,p,\dot{\d}; \v, q,\w) &+
        s(\u,p,\dot{d}; \v,q,\z)\\
      & + \left(r [P_\gamma(\d,\jump{\sigma_n})]_+, \w_n\right)_{\IN}
        = (r \f_s, \w)_{{\O}}.      
    \end{aligned}
  \end{gathered}
\end{equation*}

\paragraph{Implementation}

The described algorithms and equations have been implemented in the finite
element library \texttt{Gascoigne3d}~\cite{Gascoigne}.
We use a Cartesian finite element
mesh, which is highly refined in the region where contact takes place. We start
with a relatively coarse time-step (e.g., $\dt=2\cdot 10^{-3}$), which is
reduced sequentially, when the ball gets close to the bottom.

We use the following numerical parameters
\begin{align*}
  \gamma_C^0=1,\quad
  \sigma = 10^5,\quad 
  \alpha_d=1,\quad
  \alpha_\text{sym}=10^3.
\end{align*}
The contact relaxation parameter is chosen $\epsilon_0=\frac{1}{8}$ for the
Rubber22 test case and $\epsilon_0=\frac{1}{4}$ for PTFE6.

%% file: files/results.tex
\subsection{Results}\label{subsec:results}

In the following we shall abbreviate the methods described in
\autoref{sec.ale} as \emph{ALE}, in \autoref{sec.cutfem}
as \emph{CutFEM} and in \autoref{sec.fsi} as \emph{FSI}.

\subsubsection{PTFE6}

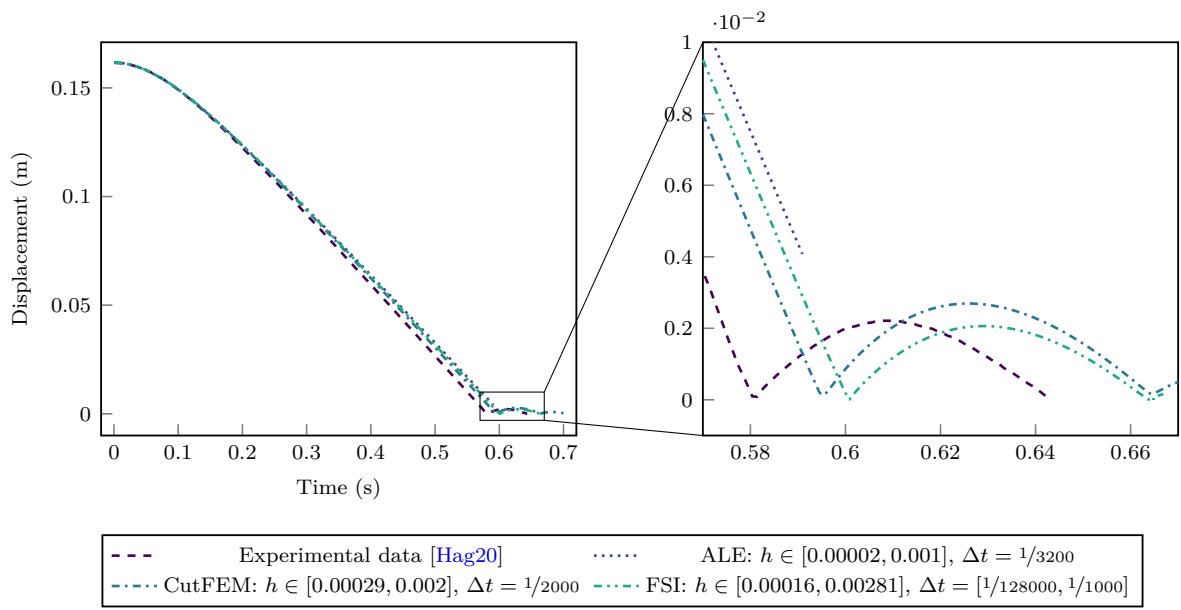
\begin{figure}
  \centering
  \input{plots/plot_results_ptfe}
  \caption{The distance between the bottom of the ball and the bottom of the
    tank: Experimental and numerical results for the PFTE6 set-up.}
  \label{fig:results:ptfe6}
\end{figure}

Our quantitative results for the PTFE6 set-up are presented in
\autoref{tab:results:PTFE6}.
\autoref{fig:results:ptfe6} shows the distance between the bottom of the ball
to the bottom of the fluid domain over time from the experimental data and the 
all three numerical methods.

If we look at the pre-contact quantities of interest in
\autoref{tab:results:PTFE6} before, we see that all methods give very
similar results for the given quantities. 
Looking at the velocity of the PTFE6 ball, we see that the
numerical values are within a relative error of $5.1\%$ of the experiment on
the finest discretisations.
Taking into account, that these results ignore the $\SI{2}{\milli\metre}$
deflection from the $z$-axis observed in the experimental data, we consider
this to be acceptable. 
Since the elastic effects of the particle appear to be negligible in this phase
of the problem and due to the known good approximation properties of the ALE
method, we consider these to be the most accurate values for future comparison.

Looking at the quantities of interest in the later phase, we see that in both
the CutFEM and FSI methods, contact occurs later than in the experiment. 
This is consistent with the smaller speed of the particle compared to the
experiments as observed above. 
With respect to the jump, we see that both methods capture the rebound
dynamics, since both the point in time at which the peak of the rebound is
realised and the size of the jump are consistent with the experiment.
As the CutFEM contact parameters were tuned with respect to the size of this
jump, this is unsurprising.
However, since the contact force only acts for a very small number of
time-steps, the fact that the time at which the rebound is maximal is also
captured well, shows that even after contact, the system is still approximated
well by the fluid-rigid body system. 
Nevertheless, it is clear that the FSI system captures the
dynamics much more accurately and without the need of essentially unknowable
parameters in the model.

\begin{table} 
  \centering
  \sisetup{
    round-mode = figures,
    round-precision = 6,
    table-format=1.5e1,
    scientific-notation=true,
    exponent-product = \cdot,
    detect-all
  }
  \resizebox{\textwidth}{!}{
  \begin{tabular}{lccrr llllll}
    \toprule
    \multicolumn{5}{c}{Discretisation} & \multicolumn{6}{c}{Results}\\
    \cmidrule(r){1-5} \cmidrule(l){6-11}
    Method & $[\hmin, \hmax]$ & $\dt$ & \texttt{dof} & \texttt{nze} & $t_\ast$
      & $v_\ast$ & $f_\ast$ & $t_\text{cont}$ & $t_\text{jump}$ & 
      $d_\text{jump}$ \\
    \midrule
    ALE
    &$[0.00008,0.004]$&\nf{1}{200 }&7.54 &0.34 &0.542134& $-0.3122077$ &
      \num{0.00111393}&--&--&--\\
    &$[0.00004,0.002]$&\nf{1}{800 }&23.88&1.10 &0.539231& $-0.3139245$ &
      \num{0.00111982}&--&--&--\\
    &$[0.00002,0.001]$&\nf{1}{3200}&83.07&3.90 &0.539026& $-0.3139931$ &
      \num{0.00112008}&--&--&--\\
    \cmidrule(lr){4-11}
    &\multicolumn{4}{r}{extrapolate}           &0.539010 & $-0.3139960$ &
      \num{0.00112021}\\
    &\multicolumn{4}{r}{order (in $h$)}&3.8  & 4.6  & 4.0\\
    \cmidrule(lr){1-11}
    CutFEM\textsuperscript{\ref{footnote:CutFEM-dofs}}
    & $[0.00114,0.008]$ & \nf{1}{2000} &  &  & 0.579181 & $-0.2884462$ &
    \num{1.1189675e-03} & 0.600328 & 0.017827 & \num{1.3397433e-03}\\
    & $[0.00057,0.004]$ & \nf{1}{2000} &  &  & 0.536698 & $-0.3164086$ &
      \num{1.1272296e-03} & 0.556037 & 0.030627 & \num{2.7068059e-03}\\
    & $[0.00029,0.002]$ & \nf{1}{2000} &  &  & 0.530537 & $-0.3213762$ &
      \num{1.1212203e-03} & 0.549543 & 0.030749 & \num{2.6924961e-03}\\
    \cmidrule(lr){1-11}
    FSI & $[0.00032, 0.00562]$ &$[\nf{1}{128\, 000}, \nf{1}{1000}]$
      &59.5 &1.24
      &0.539918 & $-0.3147132$ & -- &0.558503 &0.027515 &\num{1.81716e-03}\\
    & $[0.00016, 0.00281]$ &$[\nf{1}{128\, 000}, \nf{1}{1000}]$
      &236.6 &5.40
    &0.535625 & $-0.3159600$ & -- &0.554358 &0.028015 &\num{2.06163e-03}\\
    \cmidrule{1-11}
    \multicolumn{5}{l}{Experiment} & $0.516403$ & $-0.330987$ & -- & 
      $0.534503$ &  $0.02792$ & \num{0.00221170}\\
    \bottomrule 
  \end{tabular}
  }
  \caption{Results for the PTFE6
    set-up.\protect\textsuperscript{\ref{footnote:dof}}}
  \label{tab:results:PTFE6} 
\end{table}

\addtocounter{footnote}{1}
\footnotetext[\thefootnote]{\label{footnote:CutFEM-dofs}%
The values of \texttt{dof} and \texttt{nze} is different in
every time step, as the active level set domain changes on a constant mesh. We
therefore take rounded values after the first assembly of the system. This is
reasonable, since the mesh is quasi uniform.}

\addtocounter{footnote}{1}
\footnotetext[\thefootnote]{\label{footnote:dof}%
    \texttt{dof}: Unconstrained degrees of freedom in thousands,
    \texttt{nze}: non-zero entries of the (linearised) system in millions.}

\subsubsection{Rubber22}\label{sec.results:rubber22}

\begin{figure}
  \centering
  \input{plots/plot_results_rubber}
  \caption{The distance between the bottom of the ball and the bottom of the
    tank: Experimental and numerical results for the Rubber22 set-up.}
  \label{fig:results:rubber22}
\end{figure}
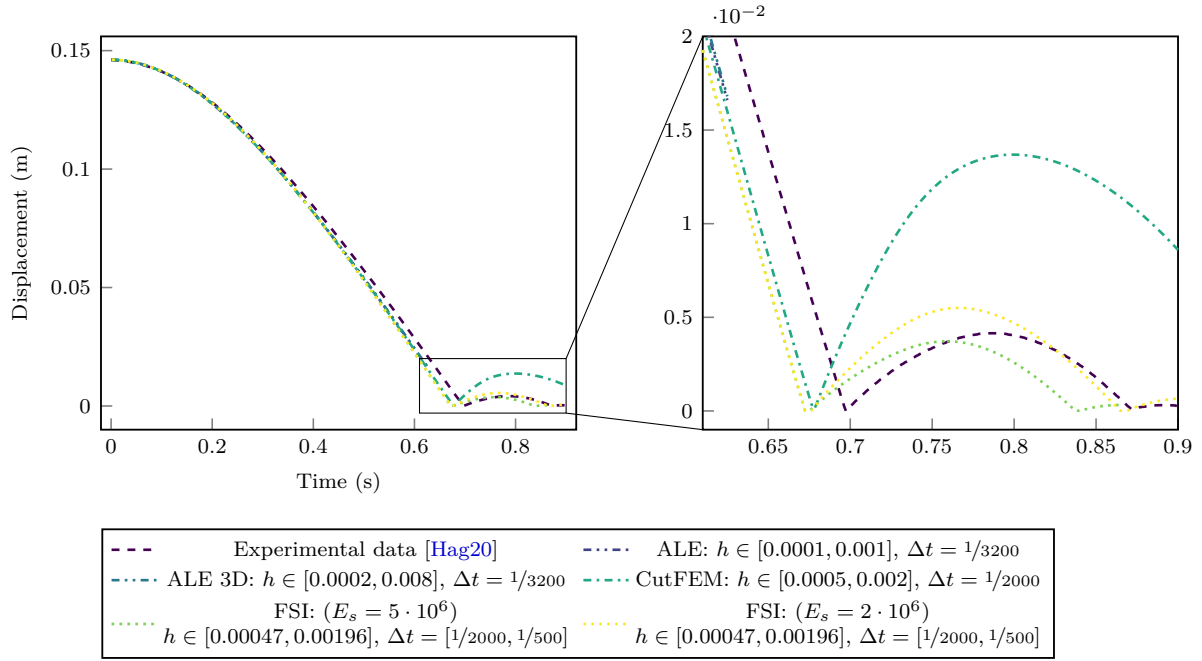

\begin{figure}
  \centering
  \includegraphics[width=0.49\textwidth]{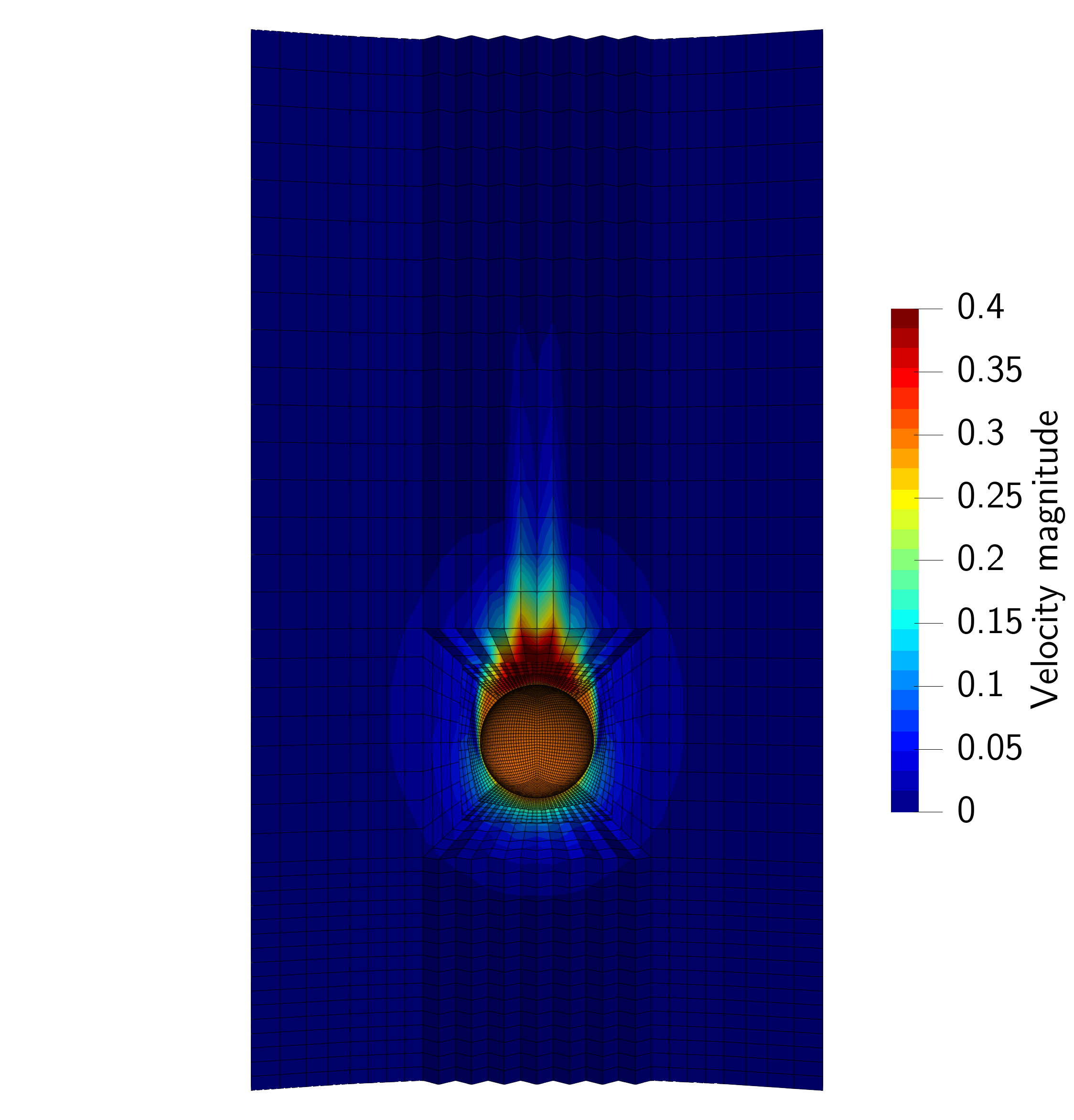}
  \includegraphics[width=0.49\textwidth]{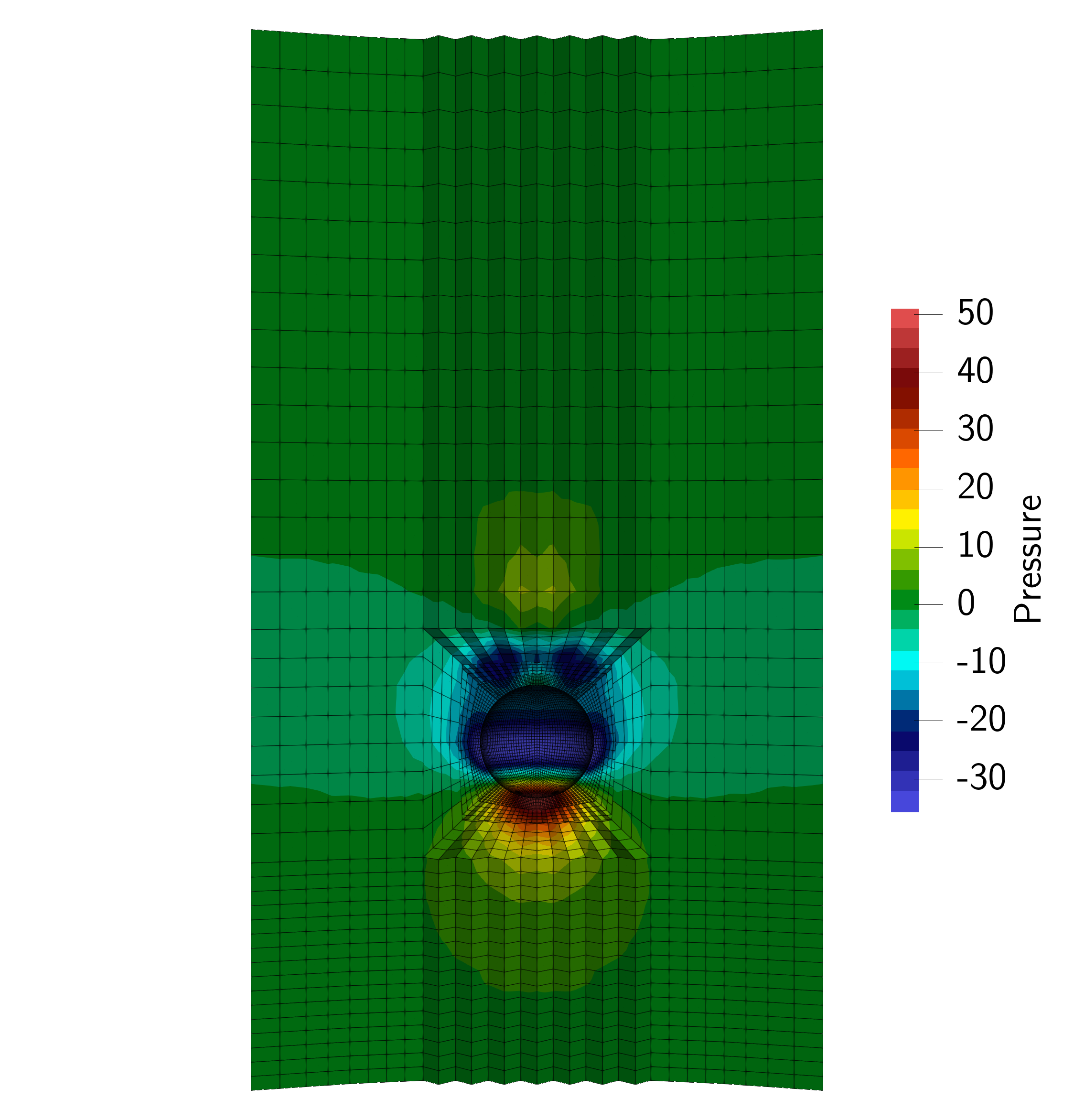}
  \caption{Velocity and pressure solution at $t=\unit[0.45]{s}$ for the
    Rubber22 case, resulting from the ALE 3D computation with
    $h_\text{max}=0.016$ and $\dt=\unit[0.005]{s}$.}
  \label{fig:results:rubber22-solution}
\end{figure}

Our results for the Rubber22 problem are presented in
\autoref{tab:results:Rubber22}. 
The height of the ball over time can be seen in \autoref{fig:results:rubber22}.
To illustrate the applicability of our spatially reduced formulation, we solve
the fluid-rigid body system in full three spatial dimensions using the ALE
approach. 
These results are also given in \autoref{tab:results:Rubber22}, while a
visualisation of the velocity and pressure solution at $t=\unit[0.45]{s}$ from
the three dimensional ALE computation can be seen in
\autoref{fig:results:rubber22-solution}.

We again start by inspecting the pre-contact results in
\autoref{tab:results:Rubber22}.
We observe that $t_\ast$, $v_\ast$ and $f_\ast$ are very similar for all
methods.
However, here the discrepancy between the numeric and experimental velocity
value is significantly smaller, with the relative difference being between
$2.1\%$ and $0.6\%$ on the finest discretisations. 
We note at this point, that the deviation from the $z$-axis in this experiment
was less than $\SI{0.75}{\milli\meter}$.
This shows that even for a more elastic material, a fluid-rigid body system can
capture the pre-contact dynamics as well as a full FSI model. 
Furthermore, we see that with perfect initial data and spatial symmetry, 
our spatially reduced model, derived under the assumption of a rotationally
symmetric solution, captures the dynamics as well as the significantly more
computationally expensive full three dimensional computation.

Looking at the numbers for the contact and rebound dynamics, we again see that
the time of contact is similar for both the CutFEM and FSI methods and this
matches the experimental time with an error of less than $5\%$. For the CutFEM
method we clearly see that the PTFE tuned parameters are not able to capture
the rebound dynamics well. In fact, the size of the rebound is approximately
$3$ times larger than the physical rebound. This shows that the use of an
artificial lubrication force, as considered in a variety of other
literature, can lead to physically meaningful results but is heavily dependent
on the ``correct'' choice of parameters for which there is no a priori
knowledge. For the FSI model, we see that the overall dynamics are
captured well. We also observe that ---while the pre-contact dynamics are
essentially independent of the elasticity parameters--- a variation of the
elasticity modulus $E_s$ changes the rebound height $d_{\text{jump}}$
significantly. Here, a softer material ($E_s = 2\cdot 10^6$ Pa) leads to a
larger rebound, as more elastic energy is taken up through the deformation
during the impact.

\begin{table} 
  \centering
  \sisetup{
    round-mode = figures,
    round-precision = 6,
    table-format=1.5e1,
    scientific-notation=true,
    exponent-product = \cdot,
    detect-all
  }
  \resizebox{\textwidth}{!}{
  \begin{tabular}{lccrr llllll}
    \toprule
    \multicolumn{5}{c}{Discretisation} & \multicolumn{6}{c}{Results}\\
    \cmidrule(r){1-5} \cmidrule(l){6-11}
    Method & $[\hmin, \hmax]$ & $\dt$ & \texttt{dof} & \texttt{nze}
      & $t_\ast$ & $v_\ast$ & $f_\ast$ & $t_\text{cont}$ & $t_\text{jump}$ & 
      $d_\text{jump}$ \\
    \midrule
    ALE
    &$[0.0004,0.004]$&\nf{1}{200} & 6.9&0.31&0.4559045 & $-0.303002$ &
      \num{0.0113262}&--&--&--\\
    &$[0.0002,0.002]$&\nf{1}{800} &21.5&0.99&0.4553551 & $-0.303552$ &
      \num{0.0113107}&--&--&--\\
    &$[0.0001,0.001]$&\nf{1}{3200}&73.8&3.45&0.4553334 & $-0.303616$ &
      \num{0.0113117}&--&--&--\\
    \cmidrule(lr){4-11}
    &\multicolumn{4}{r}{extrapolate}   &0.4553325 & $-0.303625$ &--&--&--&--\\
    &\multicolumn{4}{r}{order (in $h$)}&4.6       & 3.1         &--&--&--&--\\
    \cmidrule(lr){2-11}
    ALE 3D
    &$[0.0008,0.032]$&$\nf{1}{200}$ &18.43 &4.05&   0.453695 & $-0.304252$ &
      \num{0.01142517} & -- & -- & --\\
    &$[0.0004,0.016]$&$\nf{1}{800}$ &68.52 &15.37&  0.456145 & $-0.302170$ &
      \num{0.01144928} &  -- & -- & --\\
      &$[0.0002,0.008]$ &$\nf{1}{3200}$&304.4&69.33&0.455929 & $-0.302667$ &
      \num{0.01135569} & -- & -- & --\\
    \cmidrule(lr){1-11}
    CutFEM\textsuperscript{\ref{footnote:CutFEM-dofs}}
    & $[0.00200,0.008]$ & \nf{1}{2000} &  &  & 0.453455 & $-0.3095423$ &
      \num{1.0909363e-02} & 0.524979 & 0.098086 & \num{7.3081547e-03}\\
    & $[0.00100,0.004]$ & \nf{1}{2000} &  &  & 0.454081 & $-0.3088412$ &
      \num{1.1108214e-02} & 0.525502 & 0.110672 & \num{1.1477205e-02}\\
    & $[0.00050,0.002]$ & \nf{1}{2000} &  &  & 0.453789 & $-0.3062202$ &
      \num{1.1292175e-02} & 0.526010 & 0.121606 & \num{1.3687366e-02}\\   
    \cmidrule(lr){1-11}
    FSI ($E_s=5\cdot 10^6$) 
    & $[0.0010,0.004]$ & $[\nf{1}{2000}, \nf{1}{500}]$ & 51.4 & 1.15 & 
      0.446020 & $-0.3206889$ & -- & 0.515197 & 0.07914 &
      $2.68298\cdot10^{-3}$\\
    & $[0.0005,0.002]$ & $[\nf{1}{2000}, \nf{1}{500}]$ & 204.6 & 4.72 &
      0.449821 & $-0.3113851$ & -- & 0.521487 & 0.083323 & 
      $3.71986\cdot10^{-3}$\\
    \cmidrule(lr){2-11}
    FSI ($E_s=2\cdot 10^6$) 
    & $[0.0010, 0.004]$ & $[\nf{1}{2000}, \nf{1}{500}]$ & 51.4 & 1.15 &
      0.446012 & $-0.3207151$ & -- & 0.516197 & 0.0925 &
      $5.29247\cdot 10^{-3}$\\
    & $[0.0005, 0.002]$ & $[\nf{1}{2000}, \nf{1}{500}]$ & 204.6 & 4.72 &
      0.449827 & $-0.3113759$ & -- & 0.522087 & 0.0905 &
      $5.50146\cdot 10^{-3}$\\
    \cmidrule{1-11}
    \multicolumn{5}{l}{Experiment} & $0.469137$ & $-0.309301$ & -- & 
      $0.544021$ &  $0.089492$ & \num{0.004414849}\\
    \bottomrule 
  \end{tabular}
  }
  \caption{Results for the Rubber22 
    set-up.\protect\textsuperscript{\ref{footnote:dof}}}
  \label{tab:results:Rubber22} 
\end{table}

\subsubsection{Three dimensional computation including rotational effects}
\label{sec.ale.3d}

One of the challenges in performing the experimental study~\cite{HTR20} was
to limit the horizontal deflection of the falling objects, i.e., to keep
them close to the center line. Identifying the source of these three
dimensional effects is one of the intriguing questions for future
research. The cause may be found in a complex solution pattern of
the Navier-Stokes equations, in material inaccuracies like non-uniform
distribution of the mass or the surface roughness but also in
experimental inaccuracies, e.g.~during the release process.

The three dimensional simulations based on the ALE formulation presented
above in \autoref{sec.results:rubber22} did not show any
three dimensional effect if the configuration is fully symmetric and the
ball is released at the center line, e.g. $\c_{\SO} =
\big(0,0,d(0)\big)$. To investigate the stability of the
Navier-Stokes rigid body system we consider further numerical simulations
based on the Rubber22 case with distorted initial values.
We start the simulation with the initial data

\begin{align*}
  \c_{\SO} &=
  \big(10^{-3}\chi_x,10^{-3}\chi_y,0.1461203\big)\si{\m},\\
  \v_\SO(0) &= \big(4\cdot 10^{-3}\chi_x, 4\cdot
  10^{-3}\chi_y,0\big)\si{\m\per\s}\\
  \omegat_\SO(0) & = \big(\chi_1,\chi_2,\chi_2)~2\pi\cdot10^{-2}
    \si{\per\s},
\end{align*}
where $\chi_x,\chi_y,\chi_1,\chi_2,\chi_3\overset{\text{iid}}{\sim}
\pazocal{N}(0,1)$ are normally
distributed random numbers with mean zero and standard deviation 1.

\begin{figure}
  \centering
  \begin{minipage}{0.44\textwidth}
    \begin{flushleft}
       \includegraphics{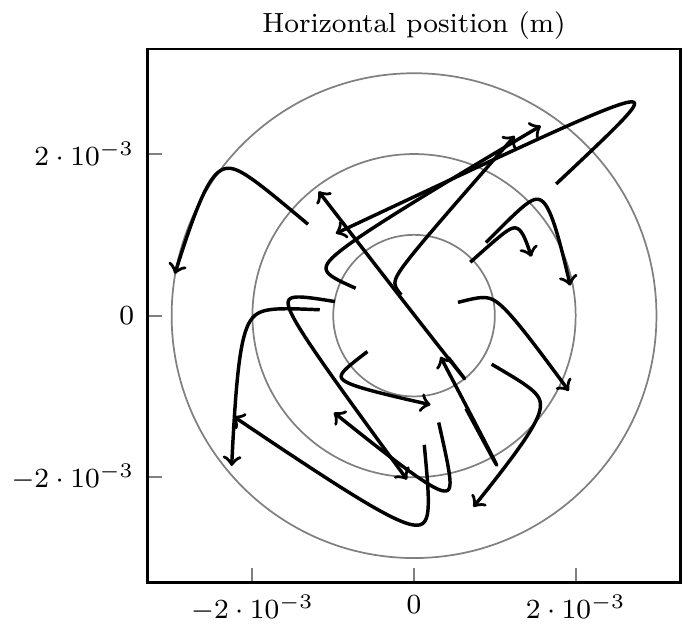}
    \end{flushleft}
  \end{minipage}
  \begin{minipage}{0.53\textwidth}
    \begin{flushright}
      \includegraphics{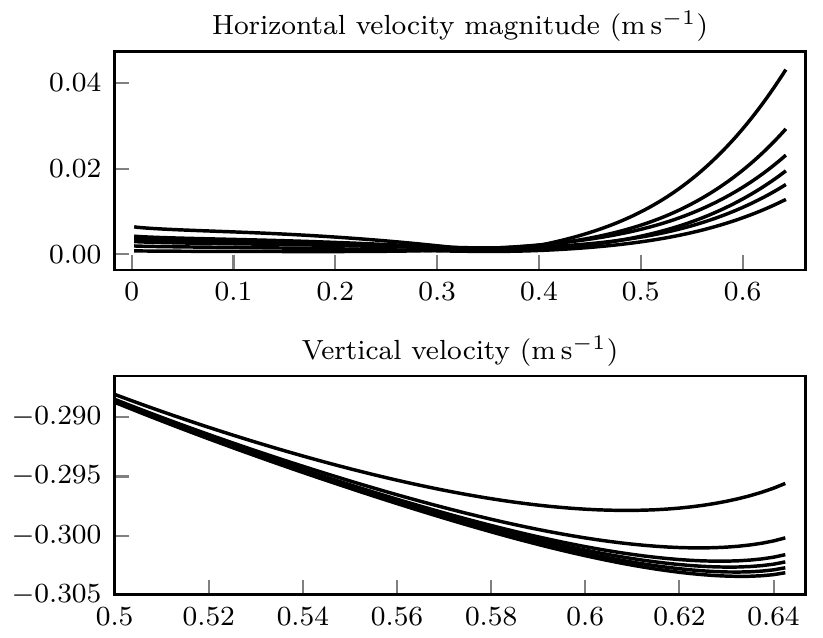}
    \end{flushright}
  \end{minipage}
  \caption{Left: View from below. Deflection of the center of mass from the
    centreline $(0,0)$ for 15 experiments starting with random
    initial deviations each. Right: velocity of the particles for 6 
    experiments. The upper figure shows the horizontal velocity component
    $\sqrt{\v_{\SO,1}^2+\v_{\SO,2}^2}$ and the lower plot gives the
    vertical velocity close to the bottom for $t\ge \unit[0.5]s$. }
  \label{3d:der}
\end{figure}

\autoref{3d:der} shows the results for multiple experiments based on
randomly chosen initial data. The left plot shows the projection of
the centre of mass onto the $x$-$y$ plane. These results show that numerical
simulations cannot predict a substantial deflection from the center
line, if an initial deflection is prescribed. While the rigid solids
are indeed further removed from the center, the effect is small and
the objects remain within $3\si{\milli\metre}$ of the center. 
On the right, we show the
velocity of the particles. The upper figure shows the horizontal
velocity component while the lower plot gives the dominant vertical
velocity. Here, we indeed see a substantial impact of the initial
disturbances. When the solid comes close to the lower boundary, a
deflection to the sides gets visible. We note that these ALE
simulations crash before contact is established. At final time
$t\approx \unit[0.64]{s}$ this distance between the lower boundary is
still slightly larger than one radius $r_\SO=\unit[0.011]{m}$. Since
the horizontal velocity is beginning to increase significantly here, 
further simulations, in which the particle is able to get closer to the bottom 
boundary, are of interest.
We deduce from these results, that small fluctuations during the release
process can indeed explain the small horizontal displacement observed during
the experiments for the PTFE and Rubber particles and the discrepancy between
the experimental and numerical realisations of our quantities of interest.
However, these computations also show, that very large horizontal
displacements, as observed in~\cite{HTR20, Hag20} for example for the POM
particle, cannot be solely explained through this.

%% file: plots/plot_results_ptfe.tex
\begin{tikzpicture}
\def\datahvwptfe{data/CutFEM/PTFE6-iso-RotSym_%
hmax0.002inner7bottom1dtinv2000BDF2_dist0.0002gc0.38.txt}
\def\datatrptfe{data/ALE/ptfe6-0.000313-83.07-3.90.txt}
\def\datasfptfe{data/FSI/ptfe_ref4.txt}
  \begin{axis}[
    width=7.9cm,
    ylabel={Displacement (m)},
    ytick={0.15, 0.1, 0.05, 0},
    yticklabels={0.15, 0.1, 0.05, 0},
    xlabel={Time (s)},
    xmin=-0.02, xmax=0.72,
    ymin=-0.01, ymax=0.171,
    restrict x to domain=0:0.7,
    legend style={at={(0,-0.25)},
                  anchor=north west,
                  /tikz/column 2/.style={column sep=3pt}},
    legend columns=2,
    ]    
    \addplot +[dashed] table [x expr=\thisrow{time}-0.1365918, y=centerY,
      col sep=tab] {data/PTFE6.txt};
    \addlegendentry{Experimental data~\cite{Hag20}}
    \addplot + [dotted] table 
      [col sep=space, header=false, x index=0, y index=1]
      {\datatrptfe};
    \addlegendentry{ALE: $h\in[0.00002, 0.001]$, $\dt=\nf{1}{3200}$}
    \addplot + [dashdotted] table [x=time, y=height, col sep=tab]
      {\datahvwptfe};
    \addlegendentry{CutFEM: $h\in[0.00029,0.002]$, $\dt=\nf{1}{2000}$}
    \addplot + [dashdotdotted] table [x index=0, y index=3, col sep=space]
      {\datasfptfe};
    \addlegendentry{FSI: $h\in[0.00016,0.00281]$, $\dt=[\nf{1}{128000},
      \nf{1}{1000}]$}
    \draw[thin]
      (axis cs:0.57,-0.003) --
      (axis cs:0.67,-0.003) coordinate (lr1) --
      (axis cs:0.67,0.01) coordinate (ur1) --
      (axis cs:0.57,0.01) -- cycle;
  \end{axis}
  \begin{axis}[
    xshift=8cm,
    width=7.9cm,
    xmin=0.57, xmax=0.67,
    ymin=-0.001, ymax=0.01,
    restrict x to domain=0.57:0.67,
    ]    
    \addplot +[dashed] table [x expr=\thisrow{time}-0.1365918, y=centerY,
      col sep=tab] {data/PTFE6.txt};
    \addplot + [dotted] table 
      [col sep=space, header=false, x index=0, y index=1]
      {\datatrptfe};
    \addplot + [dashdotted] table [x=time, y=height, col sep=tab]
      {\datahvwptfe};
    \addplot + [dashdotdotted] table [x index=0, y index=3, col sep=space]
      {\datasfptfe};
  \end{axis}
  \draw[thin]
      (current axis.north west) -- (ur1)
      (current axis.south west) -- (lr1);
\end{tikzpicture}

%% file: plots/plot_results_rubber.tex
\begin{tikzpicture}
\def\datahvwrub{data/CutFEM/Rubber22-iso-RotSym_%
hmax0.002inner4bottom1dtinv2000BDF2_dist0.0002gc12.txt}
\def\datatrrub{data/ALE/rubber22-0.000313-73.75-3.45.txt}
\def\datatrrubthr{data/ALE/rubber-3d-0.000313-304.39-69.33.txt}
\def\datasfrubfive{data/FSI/rubber22_nu04999E5e+6_ref4.txt}
\def\datasfrubtwo{data/FSI/rubber22_nu04999E2e+6_ref4.txt}
  \begin{axis}[
    width=7.9cm,
    ylabel={Displacement (m)},
    ytick={0.15, 0.1, 0.05, 0},
    yticklabels={0.15, 0.1, 0.05, 0},
    xlabel={Time (s)},
    xmin=-0.02 , xmax=0.92,
    ymin=-0.01, ymax=0.156,
    restrict x to domain=0:0.9,
    legend style={at={(0,-0.25)},
                  anchor=north west,
                  cells={align=center},
                  /tikz/column 2/.style={column sep=3pt}},
    legend columns=2,
    ]    
    \addplot +[dashed] table [x expr=\thisrow{time}-0.176659, y=centerY,
      col sep=tab] {data/Rubber22.txt};
    \addlegendentry{Experimental data~\cite{Hag20}}
    \addplot + [dashdotdotted] table [x index=0, y index=1, col sep=space]
      {\datatrrub};
    \addlegendentry{ALE: $h\in[0.0001,0.001]$, $\dt=\nf{1}{3200}$}
    \addplot + [dashdotdotted] table [x index=0, y index=3, col sep=space]
      {\datatrrubthr};
    \addlegendentry{ALE 3D: $h\in[0.0002, 0.008]$, $\dt=\nf{1}{3200}$}
    \addplot + [dashdotted] table [x=time, y=height, col sep=tab]
      {\datahvwrub};
    \addlegendentry{CutFEM: $h\in[0.0005, 0.002]$, $\dt=\nf{1}{2000}$}
    \addplot + [dotted] table [x index=0, y index=3, col sep=space]
      {\datasfrubfive};
    \addlegendentry{FSI: ($E_s = 5\cdot10^6$)\\
      $h\in[0.00047, 0.00196]$, $\dt=[\nf{1}{2000}, \nf{1}{500}]$}
    \addplot + [dotted] table [col sep=space, x index=0, y index=3]
      {\datasfrubtwo};
    \addlegendentry{FSI: ($E_s = 2\cdot10^6$)\\
      $h\in[0.00047, 0.00196]$, $\dt=[\nf{1}{2000}, \nf{1}{500}]$}
    \draw[thin]
      (axis cs:0.61,-0.003) --
      (axis cs:0.9,-0.003) coordinate (lr2) --
      (axis cs:0.9,0.02) coordinate (ur2) --
      (axis cs:0.61,0.02) -- cycle;
  \end{axis}
  \begin{axis}[
    xshift=8cm,
    width=7.9cm,
    xmin=0.61, xmax=0.9,
    ymin=-0.001, ymax=0.02,
    restrict x to domain=0.61:0.9,
    ]    
    \addplot +[dashed] table [x expr=\thisrow{time}-0.176659, y=centerY,
      col sep=tab] {data/Rubber22.txt};
    \addplot + [dashdotdotted] table [x index=0, y index=1, col sep=space]
      {\datatrrub};
    \addplot + [dashdotdotted] table [x index=0, y index=3, col sep=space]
      {\datatrrubthr};
    \addplot + [dashdotted] table [x=time, y=height, col sep=tab]
      {\datahvwrub};
    \addplot + [dotted] table [x index=0, y index=3, col sep=space]
      {\datasfrubfive};
    \addplot + [dotted] table [col sep=space, x index=0, y index=3]
      {\datasfrubtwo};
  \end{axis}
  \draw[thin]
      (current axis.north west) -- (ur2)
      (current axis.south west) -- (lr2);
\end{tikzpicture}

%% file: files/conclusion.tex
\section{Summary \& Conclusion}\label{sec:conclusion}

We have presented two set-ups for a fluid structure interaction problem with
solid contact, based on the physical experiments described in~\cite{HTR20} and
the data available at~\cite{Hag20}.
We computed these set-ups using a spatially reduced model under the assumption
of rotational symmetry in cylinder coordinates. 
For the discretisation, we used a fitted ALE and unfitted CutFEM approach
within a fluid-rigid body model and a fully Eulerian FSI approach in a
fluid-elastic structure model capable of resolving solid contact.

We showed how each of these discretisations is able to capture the pre-contact
dynamics observed in the physical experiment within a margin of $5.1\%$
to $0.6\%$, even though this ignored any horizontal motion observed in the
experimental data.
Using a full three dimensional ALE discretisation, we saw that the spatially
reduced approach did indeed result in meaningful results under the assumption
of perfect initial conditions at a fraction of the computational cost.
Furthermore, we presented computation with disturbed initial data. From these
results we deduced, that the observed horizontal motion in the PTFE 
and Rubber experiments is within the scope explainable by imperfect starting
conditions. This shows that a fluid-rigid body model is suitable for this type
of problem before solid contact occurs.

With respect to the contact dynamics, we were able to show that the Eulerian
FSI discretisation with contact treatment is able to reproduce the spatial and
temporal dynamics observed in the experiments very well. 
This is even though the theoretical modelling of such contact dynamics is
not yet fully understood.
The moving domain CutFEM approach together with the artificial contact
treatment showed that this type of contact treatment can result in physically
meaningful results when the artificial parameters are chosen ``correctly''
and the extent to which the artificial parameters are material dependent.

The resulting data sets, the source code for the fluid-rigid body
discretisations as well as some simplified examples are available in the
zenodo repository~\cite{vWRFH20}.

We conclude that the two discussed set-ups are well suited for the validation
of fluid structure interaction models in the moderate Reynolds number fluid
regime both before and after contact. 
To the best of our knowledge this is the first example of a computational FSI
set-up with rebound contact dynamics which is validated by experimental data. 
We note that the PTFE6 set-up is better suited for validation of contact and
rebound models, since the model parameters are known precisely. On the other
hand, the Rubber22 scenario is well suited to validate models before or
without contact, since there is less deviation from the centre line in the
data from the experiment.

For future research, it remains to be investigated to what extent the free
surface at the top of the fluid domain plays a role in system dynamics.
Furthermore, an open question is that of the role of imperfections in the
mass distribution within the solid, of the surface roughness of the solid
and whether these can explain larger horizontal displacements and rotation
observed for example with the POM particle in~\cite{HTR20}.

%% file: files/appendix.tex
\section{Computational test cases}\label{appendix}

We define two simplified test cases. This is intended make it easier for others
to reproduce the presented results using different methods and/or
implementations.

\subsection{Stationary flow test}

For this stationary test, we modify the Rubber22 set up. The sphere is fixed at
$c_\SO=(0,0,0.1)$, i.e., the centre of the cylinder. We impose an inflow
boundary condition $\u=-0.01(1 - (\x_1^2 + x_2^2)/R^2)$ on $\Gtop$, no-slip
$\u=0$ on $\Gwall\cup\IN$ and a homogeneous Neumann boundary condition
$\stress\n=\bm{0}$ on $\Gbot$. 

We consider the stationary Navier-Stokes problem on this domain. As reference
quantities we take the vertical stress acting on the sphere, i.e.,
testing the reduced formulation in \eqref{eqn:rotated-stress} with the
non-conforming, continuous test-functions $\widehat\w = (0, 1)^T$ on $\IN$ and 
$\bm{0}$ on $\G$ respectively. 

We compute the problem based on the discretisations discussed in
\autoref{sec.ale} and \autoref{sec.cutfem}.
The results can be seen in \autoref{tab:test1}.

\begin{table}[!htp]
  \sisetup{
    round-mode = figures,
    round-precision = 6,
    table-format=1.5e1,
    scientific-notation=true,
    exponent-product = \cdot,
    detect-all
  }
  \centering
  \begin{tabular}{llrr l}
    \toprule
    \multicolumn{4}{c}{Discretisation} & Results\\
    \cmidrule(r){1-4}\cmidrule(l){5-5}
    Method & $[\hmin, \hmax]$ & \texttt{dof} & \texttt{nze} & $\bm{F}_z$\\
    \midrule
    ALE
    &$[0.00030, 0.0049]$& 6.9  & 0.310 & \num{-4.432338e-05} \\
    &$[0.00015, 0.0024]$& 26.7 & 1.238 & \num{-4.429920e-05} \\
    &$[0.00008, 0.0012]$& 104.9& 4.946 & \num{-4.429749e-05} \\
    \cmidrule(lr){3-5}
    & & \multicolumn{2}{r}{extrapolate} & \num{-0.000044297364}\\
    & & \multicolumn{2}{r}{order (in h)} & 3.8 \\
    \cmidrule(lr){1-5}
    CutFEM
    & $[0.0020,0.008]$ & 11.9 & 0.361 & \num{-4.40253989e-05}\\
    & $[0.0010,0.004]$ & 41.3 & 1.241 & \num{-4.42354979e-05}\\
    & $[0.0005,0.002]$ & 151.4 & 4.525 & \num{-4.42918828e-05}\\
    \bottomrule
  \end{tabular}
  \caption{Resulting reference quantities for the stationary test 
    scenario.\protect\textsuperscript{\ref{footnote:dof}}}
  \label{tab:test1}
\end{table}

\subsection{Non-stationary flow test with prescribed motion}

As a second test case we keep the Rubber22 set-up as the basis. The material
parameters and the cylinder boundary conditions are as before, i.e.,
we consider no-slip on $\Gwall\cup\Gbot$ and a free-slip condition on $\Gtop$.
We prescribe the motion of the sphere as follows. Over the time-interval $I=[0,
20]$, the sphere is located at
\begin{equation*}
  c_\SO = \big(0, 0, d(t)\big)\quad\text{with}\quad d(t) = 0.1 +
    0.05\cos(0.1\pi t).
\end{equation*}
The boundary condition on the interface is accordingly set to $\u=\big(0,
0,\partial_t d(t)\big)$. Quantities of interest are the maximal value
over time of the $z$-component of the force functional in the reduced
formulation. 

\begin{figure}[!htp]
  \centering
  \input{plots/plot_test_force}
  \caption{Force functionals acting on the sphere with prescribed motion 
    over time.}
  \label{fig:test2}
\end{figure}
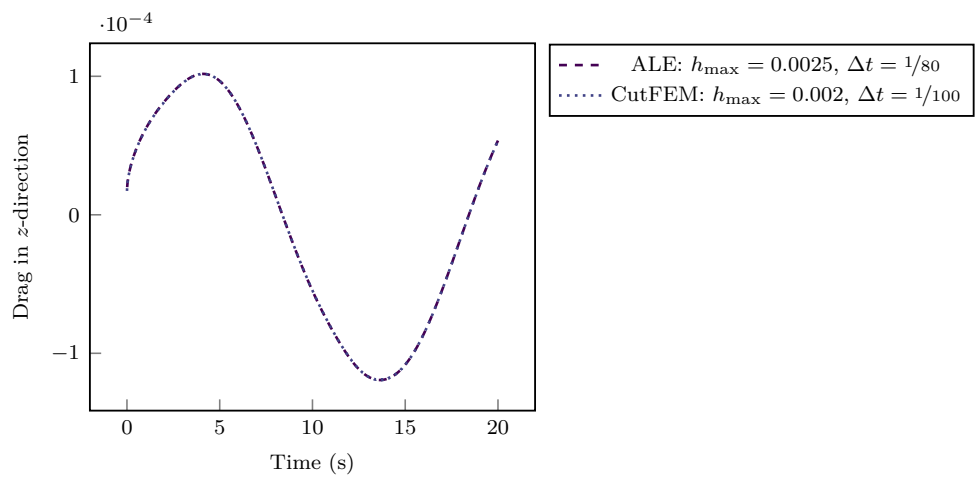

We compute this again using the reduced formulation with the rigid body
discretisations. The quantitative results can be seen in \autoref{tab:test2}
while the force functional is shown over time in \autoref{fig:test2}.

\begin{table}[!htp]
  \sisetup{
    round-mode = figures,
    round-precision = 6,
    table-format=1.5e1,
    scientific-notation=true,
    exponent-product = \cdot,
    detect-all
  }
  \centering
  \begin{tabular}{lllrr ll}
    \toprule
    \multicolumn{5}{c}{Discretisation} & \multicolumn{2}{c}{Results}\\
    \cmidrule(r){1-5}\cmidrule(l){6-7}
    Method & $[\hmin,\hmax]$ & $\dt$ & \texttt{dof} & \texttt{nze} &
      $\bm{F}_{z,\text{max}}$ & $t_{z,\text{max}}$\\
    \midrule
    ALE
    &$[0.00030,0.0099$]& $\nf{1}{5}$  & 11.1 & 0.49 & \num{1.01837834e-04} &
      4.10734691\\
    &$[0.00015,0.0050]$& $\nf{1}{20}$ & 25.0 & 1.10 & \num{1.01748104e-04} &
      4.11356325\\
    &$[0.00008,0.0025]$& $\nf{1}{80}$ & 58.7 & 2.61 & \num{1.01720245e-04} &
      4.10670436\\
    \cmidrule(lr){1-7}
    CutFEM\textsuperscript{\ref{footnote:CutFEM-dofs}}
    & $[0.0020,0.008]$ & \nf{1}{25} & 11.9 & 0.361 & \num{1.016999e-04} &
      4.153\\
    & $[0.0010,0.004]$ & \nf{1}{50} & 41.2 & 1.240 & \num{1.013946e-04} &
      4.138\\
    & $[0.0005,0.002]$ & \nf{1}{100} & 151.3 & 4.526 & \num{1.016712e-04} &
      4.111\\
    \bottomrule
  \end{tabular}
  \caption{Resulting reference quantities for the non-stationary moving domain
    test scenario.\protect\textsuperscript{\ref{footnote:dof}}}
  \label{tab:test2}
\end{table}

%% file: plots/plot_test_force.tex
\begin{tikzpicture}
\def\datahw{data/CutFEM/Output_TimeDepTest_diam0.022%
mu0.008rho1141_iso-hmax0.002dtinv100.0BDF2.txt}
\def\datatr{data/ALE/moving-rubber22-0.0125-58671.txt}
  \begin{axis}[
    width=7.5cm,
    ylabel={Drag in $z$-direction},
    xlabel={Time (s)},
    legend pos = outer north east,
    legend style={cells={align=left}},
    legend columns=1,
  ]
    \addplot +[dashed] table 
      [col sep=space, header=false, x index=0, y index=4] {\datatr};
    \addlegendentry{ALE: $\hmax=0.0025$, $\dt=\nf{1}{80}$}
    \addplot +[dotted] table [x=time, y=drag_z, col sep=tab] {\datahw};
    \addlegendentry{CutFEM: $\hmax=0.002$, $\dt=\nf{1}{100}$}
  \end{axis}
\end{tikzpicture}